\documentclass[final,5p,times,twocolumn]{elsarticle}
\usepackage{amssymb}
\usepackage{multirow}
\usepackage{lscape}
\usepackage{amsmath}
\usepackage{pdfpages}

\journal{Nuclear Physics B}

\begin{document}

\begin{frontmatter}



\title{Complementary Recommendation in E-commerce: Definition, Approaches, and Future Directions}


\author[label1]{Linyue Li}

\author[label1,label2]{Zhijuan Du\corref{cor1}}
\ead{nmg-duzhijuan@163.com}

\address[label1]{Inner Mongolia University, Hohhot, China}
\address[label2]{Engineering Research Center of Ecological Big Data, Ministry of Education, China}
\cortext[cor1]{Corresponding author}
\begin{abstract}
In recent years, complementary recommendation has received extensive attention in the e-commerce domain. In this paper, we comprehensively summarize and compare 34 representative studies conducted between 2009 and 2024. Firstly, we compare the data and methods used for modeling complementary relationships between products, including simple complementarity and more complex scenarios such as asymmetric complementarity, the coexistence of substitution and complementarity relationships between products, and varying degrees of complementarity between different pairs of products. Next, we classify and compare the models based on the research problems of complementary recommendation, such as diversity, personalization, and cold-start. Furthermore, we provide a comparative analysis of experimental results from different studies conducted on the same dataset, which helps identify the strengths and weaknesses of the research. Compared to previous surveys, this paper provides a more updated and comprehensive summary of the research, discusses future research directions, and contributes to the advancement of this field.
\end{abstract}

\begin{keyword}
	Recommender system \sep Complementary recommendation \sep E-commerce \sep Artificial intelligence \sep Survey
\end{keyword}

\end{frontmatter}

\section{Introduction}
In today's digital era, recommendation systems have become a key technology in many industries and have received widespread attention from a large number of domestic and foreign researchers. The main goal of traditional recommendation systems is to recommend similar content or products to users based on their historical behavior and interests. Techniques such as collaborative filtering, content filtering and hybrid recommendation are mainly used. However, similar recommendations may not necessarily meet the needs of users and merchants. For users, traditional recommendations will create recommendation redundancy. For example, if a user has just purchased a laptop, it would be meaningless to recommend a laptop to the user at this time. For merchants, if similar products are always recommended, the user purchase rate will decrease, which will affect the merchant's sales. Therefore, in order to solve the problem of recommendation homogeneity, research on complementary recommendations has received increasing attention.
\newline \indent Complementary recommendations refer to recommending other products to users that can be used in conjunction with the products they purchased or browsed. For example, if a user purchases a laptop, related accessories can be recommended to the user, such as mice, keyboards, external monitors, headphones, etc; If a user purchases a skincare product, other products such as cleansers, masks, serums or sunscreens can be recommended to fulfill the user's complete skincare regimen. These two examples embody the idea of complementary recommendations. Compared to traditional recommendations, complementary recommendations help users discover products that may have been overlooked, improving the user shopping experience and merchant sales.
\newline \indent However, the existing complementary recommendation review work is not perfect, and it lacks the introduction and comparative analysis of the latest technologies in recent years. In order to make up for the shortcomings of current work, this paper reviews complementary recommendation from the aspects of complementary definition, complementary relationship modeling, complementary recommendation models and experimental results.
\newline \indent The main structure of this article is arranged as follows:
\begin{itemize}
	\item The first chapter mainly introduces the definition of complementary relationship, application scenarios, research significance, the current situation, and contribution.
	\item Chapter 2 is to model the complementary relationship between products, which is divided into simple complementary relationship (Section 2.1) and complex complementary relationship (2.2).
	\item Chapter 3 introduces complementary recommendation models, which are classified according to research topics, mainly divided into diversity, personalization, cold start, scenario-based recommendation and data noise.
	\item Chapter 4 introduces the experimental results, which are divided into datasets, model training and evaluation metrics.
	\item Finally, Chapter 5 introduces future research and prospects, and Chapter 6 is the conclusion of this article.
\end{itemize}
\subsection{complementary relationship definition}
Complementary relationship generally refers to the interdependent and mutually complementary relationship between two things or concepts. In a complementary relationship, the existence or characteristics of one thing or concept need to be combined with another thing or concept to form a complete or effective whole. In recommendation systems, complementary relationships refer to the existence of a mutually complementary relationship between two or more items, that is, they can be used together to enhance the user experience to some extent. As shown in Table \ref{tab:tab0}, complementary relationships usually have the following forms:
\begin{table*}[hbt]
	\centering
	\caption{complementary definition}
	\label{tab:tab0}
	\resizebox{\textwidth}{!}{%
		\begin{tabular}{|c|c|c|c|c|}
			\hline
			literature & definition & carve out ideas & advantage & shortcoming \\ \hline
			\begin{tabular}[c]{@{}c@{}}\cite{DBLP:conf/www/BibasSJ23,yan2022personalized,zhao2017recommending,mcauley2015inferring,DBLP:conf/recsys/Papso23,DBLP:conf/cikm/ZhouWHZMD22,liu2020decoupled,liu2021item}\\ \cite{	hao2020p,rakesh2019linked,wang2018path,zhang2019identifying,zhang2018quality,wan2018representing,trofimov2018inferring}\end{tabular} &buy together & frequency of buy together  &simple way to judge & \begin{tabular}[c]{@{}c@{}}buy together may not\\ necessarily complement\end{tabular} \\ \hline
			~\cite{DBLP:conf/aaai/ChenHXFLSYQ23,zhang2021learning}&use together &\begin{tabular}[c]{@{}c@{}}products are used together\\  to fulfill specific needs\end{tabular}  & be in line with the situation & complex collocation \\ \hline
			~\cite{zhao2017deep,kang2019complete}&collocation compatibility& clothing style, material& suitable for fashion field & hard to measure \\ \hline
		\end{tabular}%
	}
\end{table*}
\newline \indent \textbf{Buy together:} 
It is mainly portrayed on the basis of the frequency of co-purchase of products; the more often two products are purchased together the stronger the degree of complementarity. Co-purchase is divided into two cases, the first is the user in the same order under the common purchase of product pairs, but taking into account the actual user ordering habits, the number of the first case is relatively small. So the second situation is to set a time period (such as 10 minutes), in this time period the same user to order the purchase of goods even if the complementary products.
This definition of judgment is simple, but only applies to a small number of products, because co-purchased products are not necessarily complementary, for example, the user may buy two different brands of paper towels at the same time, so the direct use of co-purchase data to determine the complementary relationship between the commodities will produce data noise problems.
\newline \indent\textbf{Use together:}
The two products have the same usage scenario and work together for better overall utility.For example, a charger needs to be used with a cellphone to make sense.
This way of defining is more realistic, but the way products are paired in daily life is more complicated and there is no ready dataset. So it is better to use the  co-purchase data of products to represent the use of products in collocation.
\newline \indent\textbf{Collocation compatibility:} 
It is mainly used for clothing complementary recommendations. Characterize according to the style, material, size, etc. of the clothing. But compatibility between outfits is difficult to measure.
\subsection{Common scenarios}
Complementary recommendation can be applied to a variety of different scenarios. The following are some common complementary recommendation scenarios:
\newline \indent\textbf{Electronic business platform:} On e-commerce platforms, complementary products related to their purchase history and browsing behavior can be recommended to users. For example, when a user purchases a mobile phone, the platform can recommend related products such as mobile phone films, mobile phone cases, and chargers. Such complementary recommendations can increase users' purchase intention and order value.
\newline \indent\textbf{Sports app:} In sports apps, users can be recommended recipes or healthy products that match their exercise plans. It can help users achieve better fitness results.
\newline \indent\textbf{Travel and hotel booking platform:} On travel and hotel booking platforms, users can be recommended complementary services related to their travel destinations and preferences. For example, when a user books a hotel, the platform can recommend nearby attractions, restaurants, car rental services, etc. to provide a more complete travel experience.
\newline \indent\textbf{Takeaway platform: }On takeaway platforms, users can be recommended complementary dishes or drinks related to the food they ordered. This helps users discover new food options while increasing order value and user satisfaction.
\subsection{Research meaning}
Complementary recommendation algorithms have important research significance for merchants, users, and platforms.
\subsubsection{Merchant aspect}
\
\newline \indent\textbf{Increase sales: }When a user purchases a product, the merchant can recommend other products that complement the product and guide the user to make additional purchases, thereby increasing sales and order value.
\newline \indent\textbf{Market analysiss: }Merchants can analyze the click-through rate, conversion rate and other indicators of complementary recommendations to understand users' cross-buying patterns and preferences, thereby conducting market analysis. It can help merchants adjust product portfolios, locate market demand, and optimize sales strategies and marketing activities.
\subsubsection{User aspect}
\
\newline \indent\textbf{Increase shopping convenience: }Through complementary recommendations, users can more easily find other products they may need. This avoids the hassle of users having to search individually or browse multiple pages to find related items. Allowing users to enjoy a better shopping experience.
\newline \indent\textbf{Broaden interests: }Complementary recommendations can help users explore and develop new areas or product categories that may be of interest to them. By recommending items that are related to, but slightly different from, their purchasing behavior, users have the opportunity to try new products and experiences, broaden their interests, and enrich their lives.
\subsubsection{Platform aspect}
\
\newline \indent\textbf{Improve user stickiness: }Complementary recommendations can increase users' stay time on the platform and frequency of interaction. By showing users relevant recommended content, platforms can provide a more engaging user experience, making users more inclined to browse, purchase or interact more on the platform. This helps improve user stickiness and increase user retention and loyalty.
\newline \indent\textbf{Improve user stickiness: }Complementary recommendations can enhance the competitiveness and differentiation advantages of the platform. By providing personalized and accurate complementary recommendations, the platform can meet the diverse needs of users and provide a better user experience, thereby attracting more users and merchants to choose the platform. This helps strengthen the platform's position in the market, improve market share and competitive advantage.
\newline \indent Overall, research on complementary recommendation algorithms not only has a positive impact on user experience, but also creates more opportunities and potential for merchants and platforms.
\subsection{State of the complementary recommendation}
The current status of complementary recommendation can be summarized as follows:
\newline \indent \textbf{Data Sources and Modeling Methods.} Currently, complementary recommendation primarily relies on publicly available e-commerce platform data, such as Amazon. The modeling methods mainly utilize machine learning and neural network technologies, such as GNN (Graph Neural Networks), MLP (Multilayer Perceptrons), and others.
\newline \indent \textbf{Focus and Research Directions.} Current research primarily focuses on modeling the complementary relationships between products while also making recommendations personalized and diverse to enhance user satisfaction. Researchers are interested in how to discover complementary relationships from users' purchase histories and how to leverage these relationships to optimize recommendation results.
\newline \indent \textbf{Challenges and Limitations.} Although complementary recommendation can improve recommendation accuracy to a certain extent, it still faces challenges and limitations. These include data sparsity, singularity of recommendation results, interpretability of the model, among others, which impact the accuracy and practicality of complementary recommendation.
\subsection{Contribution}
Currently, to our knowledge, there is only one review article in the field of complementary recommendation published in 2019 \citep{yu2019complementary}. This article investigated 23 papers published between 1994 and 2018, while our paper includes a significant amount of recent research. Regarding the modeling of complementary relationships, \citep{yu2019complementary} merely classified the papers without providing detailed descriptions of the specific data and implementation methods used. In terms of introducing complementary recommendation models, \citep{yu2019complementary} only provided model descriptions without mentioning the challenges encountered by the models or the problems they aim to solve. In terms of evaluation metrics, \citep{yu2019complementary} did not compare the experimental results of different papers or analyze the effectiveness of the models. Lastly, \citep{yu2019complementary} did not provide suitable research directions. Given these limitations, our research addresses these issues and can help researchers gain a better understanding of complementary recommendation.
\section{Modeling complementary Representation}
\subsection{Simple complementary relationship}
As shown in the Table \ref{table2}, product complementary relationship learning methods are divided into three categories based on different feature data types: (1) Learning complementary relationships based on product content; (2) Learning complementary relationships based on user purchase sequences; (3) Learn complementary relationships based on product relationship graphs
\subsubsection{Learning complementary relationships based on product content}
\
\newline Learning product embeddings for relationship prediction based on product content.
\newline \indent \textbf{Learning complementary relationships based on text content. }In terms of feature extraction,\citep{mcauley2015inferring,zhang2019identifying}uses the LDA model to obtain feature representation based on product review information. However, most review texts are very short and lack contextual information, and LDA has poor performance on short texts. \citep{rakesh2019linked,zhao2017deep,angelovska2021siamese}use neural networks, of which \citep{rakesh2019linked}uses Variational Autoencoders (VAE) to obtain feature representation based on product content; \citep{zhao2017deep}uses Convolutional Neural Network(CNN) to get feature representations based on product titles.
Since CNN are unable to capture the dependencies between words when dealing with product titles, \citep{angelovska2021siamese}improves on the basis of \citep{zhao2017deep}and gets better results by using Long Short-Term Memory(LSTM) for feature extraction of product titles.
\newline \indent When predicting the complementary relationship between products, \citep{mcauley2015inferring}uses Logistic Regression(LR) to convert the complementary relationship between products into a binary classification problem; \citep{zhang2019identifying}proposed the formula (\ref{eqn:30})-(\ref{eqn:303}) to capture the relationship between products, add 7 non-text features, and finally use the softmax activation function to obtain the final relationship prediction result.
\begin{equation}
	\phi_{1}(i, j)  =\theta_{i, k} \times \theta_{j, l}, \forall k, l \in[1, K]\label{eqn:30}
\end{equation}
\begin{equation}
	\phi_{2}(i, j)  =\left(1-\theta_{i, k}\right) \times \theta_{j, l}, \forall k, l \in[1, K]\label{eqn:301}
\end{equation}
\begin{equation}
	\phi_{3}(i, j)  =\theta_{i, k} \times\left(1-\theta_{j, l}\right), \forall k, l \in[1, K]\label{eqn:302}
\end{equation}
\begin{equation}
	\phi_{4}(i, j)  =\left(1-\theta_{i, k}\right) \times\left(1-\theta_{j, l}\right), \forall k, l \in[1, K]\label{eqn:303}
\end{equation}
\begin{landscape}
	\begin{table}[htb]
		\centering
		\caption{Classification of feature extraction methods in complementary recommendation}
		\label{table2}
		\resizebox{1.3\textwidth}{2in}{
			\begin{tabular}{|c|ccccccccccc|ccccccccccc|}
				\hline
				\multirow{3}{*}{literature} & \multicolumn{11}{c|}{data type}                                                                                                                                                                                                                                                                                                             & \multicolumn{11}{c|}{feature extraction}                                                                                                                                                                                                                                                                                                                                \\ \cline{2-23} 
				& \multicolumn{1}{c|}{\multirow{2}{*}{image}} & \multicolumn{5}{c|}{text}                                                                        & \multicolumn{2}{c|}{User purchase sequence}& \multicolumn{3}{c|}{goods relationship graph}& \multicolumn{1}{c|}{\multirow{2}{*}{MLP}} &\multicolumn{1}{c|}{\multirow{2}{*}{Prod2vec}} & \multicolumn{1}{c|}{\multirow{2}{*}{triple2vec}} & \multicolumn{1}{c|}{\multirow{2}{*}{ResNet}} & \multicolumn{1}{c|}{\multirow{2}{*}{CNN}} & \multicolumn{1}{c|}{\multirow{2}{*}{LSTM}} & \multicolumn{1}{c|}{\multirow{2}{*}{LDA}} & \multicolumn{1}{c|}{\multirow{2}{*}{GCN}} & \multicolumn{1}{c|}{\multirow{2}{*}{GAT}} & \multicolumn{1}{c|}{\multirow{2}{*}{Transformer}} & \multirow{2}{*}{others} \\ \cline{3-12}
				& \multicolumn{1}{c|}{}                    & \multicolumn{1}{c|}{title} & \multicolumn{1}{c|}{review} & \multicolumn{1}{c|}{category} & \multicolumn{1}{c|}{brand} & \multicolumn{1}{c|}{others} & \multicolumn{1}{c|}{Complete sequence} & \multicolumn{1}{c|}{\textless{}item,item,user\textgreater{}} & \multicolumn{1}{c|}{\begin{tabular}[c]{@{}c@{}}heterogeneous\\  graph\end{tabular}} & \multicolumn{1}{c|}{\begin{tabular}[c]{@{}c@{}}knowledge\\ graph\end{tabular}} & \begin{tabular}[c]{@{}c@{}}complementary\\ graph\end{tabular}  & \multicolumn{1}{c|}{}                     & \multicolumn{1}{c|}{}                          & \multicolumn{1}{c|}{}                            & \multicolumn{1}{c|}{}                        & \multicolumn{1}{c|}{}                     & \multicolumn{1}{c|}{}                      & \multicolumn{1}{c|}{}       & \multicolumn{1}{c|}{}   & \multicolumn{1}{c|}{}   & \multicolumn{1}{c|}{}                 &                      \\ \hline
				
				\citep{DBLP:conf/aaai/ChenHXFLSYQ23}  & \multicolumn{1}{c|}{}                    & \multicolumn{1}{c|}{}   & \multicolumn{1}{c|}{}   & \multicolumn{1}{c|}{}   & \multicolumn{1}{c|}{}   & \multicolumn{1}{c|}{}   & \multicolumn{1}{c|}{}       & \multicolumn{1}{c|}{}                                     & \multicolumn{1}{c|}{$\checkmark$}      & \multicolumn{1}{c|}{}      &       & \multicolumn{1}{c|}{}                     & \multicolumn{1}{c|}{$\checkmark$}                          & \multicolumn{1}{c|}{}                            & \multicolumn{1}{c|}{}                        & \multicolumn{1}{c|}{}                     & \multicolumn{1}{c|}{}                      & \multicolumn{1}{c|}{}     & \multicolumn{1}{c|}{$\checkmark$}   & \multicolumn{1}{c|}{}   & \multicolumn{1}{c|}{}                   &                      \\ \hline
				
				\citep{DBLP:conf/www/BibasSJ23}& \multicolumn{1}{c|}{$\checkmark$}                    & \multicolumn{1}{c|}{}   & \multicolumn{1}{c|}{}   & \multicolumn{1}{c|}{$\checkmark$}   & \multicolumn{1}{c|}{}   & \multicolumn{1}{c|}{$\checkmark$}   & \multicolumn{1}{c|}{}       & \multicolumn{1}{c|}{}                                     & \multicolumn{1}{c|}{}      & \multicolumn{1}{c|}{}      &       & \multicolumn{1}{c|}{$\checkmark$}                     & \multicolumn{1}{c|}{}                          & \multicolumn{1}{c|}{}                            & \multicolumn{1}{c|}{}                        & \multicolumn{1}{c|}{$\checkmark$}                     & \multicolumn{1}{c|}{}                      & \multicolumn{1}{c|}{}                 & \multicolumn{1}{c|}{}   & \multicolumn{1}{c|}{}   & \multicolumn{1}{c|}{}       &                      \\ \hline
				
				\citep{yan2022personalized}& \multicolumn{1}{c|}{}                    & \multicolumn{1}{c|}{}   & \multicolumn{1}{c|}{}   & \multicolumn{1}{c|}{}   & \multicolumn{1}{c|}{}   & \multicolumn{1}{c|}{}   & \multicolumn{1}{c|}{$\checkmark$}       & \multicolumn{1}{c|}{}                                     & \multicolumn{1}{c|}{}      & \multicolumn{1}{c|}{}      & $\checkmark$    & \multicolumn{1}{c|}{$\checkmark$}                     & \multicolumn{1}{c|}{}                          & \multicolumn{1}{c|}{}                            & \multicolumn{1}{c|}{}                        & \multicolumn{1}{c|}{}                     & \multicolumn{1}{c|}{}                      & \multicolumn{1}{c|}{}       & \multicolumn{1}{c|}{}   & \multicolumn{1}{c|}{$\checkmark$}   & \multicolumn{1}{c|}{$\checkmark$}                 &                      \\ \hline
				
				\citep{zhang2021learning}& \multicolumn{1}{c|}{}                    & \multicolumn{1}{c|}{}   & \multicolumn{1}{c|}{}   & \multicolumn{1}{c|}{}   & \multicolumn{1}{c|}{}   & \multicolumn{1}{c|}{}   & \multicolumn{1}{c|}{$\checkmark$}       & \multicolumn{1}{c|}{}                                     & \multicolumn{1}{c|}{$\checkmark$}      & \multicolumn{1}{c|}{}      &       & \multicolumn{1}{c|}{}                     & \multicolumn{1}{c|}{}                          & \multicolumn{1}{c|}{}                            & \multicolumn{1}{c|}{}                        & \multicolumn{1}{c|}{}                     & \multicolumn{1}{c|}{}                      & \multicolumn{1}{c|}{}              & \multicolumn{1}{c|}{$\checkmark$}   & \multicolumn{1}{c|}{}   & \multicolumn{1}{c|}{$\checkmark$}          &     $\checkmark$                \\ \hline
				
				\citep{zhao2017deep}& \multicolumn{1}{c|}{}                    & \multicolumn{1}{c|}{$\checkmark$}   & \multicolumn{1}{c|}{}   & \multicolumn{1}{c|}{}   & \multicolumn{1}{c|}{}   & \multicolumn{1}{c|}{}   & \multicolumn{1}{c|}{}       & \multicolumn{1}{c|}{}                                     & \multicolumn{1}{c|}{}      & \multicolumn{1}{c|}{}      &       & \multicolumn{1}{c|}{}                     & \multicolumn{1}{c|}{}                          & \multicolumn{1}{c|}{}                            & \multicolumn{1}{c|}{}                        & \multicolumn{1}{c|}{$\checkmark$}                     & \multicolumn{1}{c|}{}                      & \multicolumn{1}{c|}{}      & \multicolumn{1}{c|}{}   & \multicolumn{1}{c|}{}   & \multicolumn{1}{c|}{}                  &                      \\ \hline
				
				\citep{zhao2017recommending}& \multicolumn{1}{c|}{}                    & \multicolumn{1}{c|}{}   & \multicolumn{1}{c|}{}   & \multicolumn{1}{c|}{}   & \multicolumn{1}{c|}{}   & \multicolumn{1}{c|}{$\checkmark$}   & \multicolumn{1}{c|}{}       & \multicolumn{1}{c|}{}                                     & \multicolumn{1}{c|}{}      & \multicolumn{1}{c|}{}      &       & \multicolumn{1}{c|}{}                     & \multicolumn{1}{c|}{}                          & \multicolumn{1}{c|}{}                            & \multicolumn{1}{c|}{}                        & \multicolumn{1}{c|}{}                     & \multicolumn{1}{c|}{}                      & \multicolumn{1}{c|}{}      & \multicolumn{1}{c|}{}   & \multicolumn{1}{c|}{}   & \multicolumn{1}{c|}{}                  &       $\checkmark$              \\ \hline
				
				\citep{mcauley2015inferring}& \multicolumn{1}{c|}{}                    & \multicolumn{1}{c|}{}   & \multicolumn{1}{c|}{$\checkmark$}   & \multicolumn{1}{c|}{}   & \multicolumn{1}{c|}{}   & \multicolumn{1}{c|}{}   & \multicolumn{1}{c|}{}       & \multicolumn{1}{c|}{}                                     & \multicolumn{1}{c|}{}      & \multicolumn{1}{c|}{}      &       & \multicolumn{1}{c|}{}                     & \multicolumn{1}{c|}{}                          & \multicolumn{1}{c|}{}                            & \multicolumn{1}{c|}{}                        & \multicolumn{1}{c|}{}                     & \multicolumn{1}{c|}{}                      & \multicolumn{1}{c|}{$\checkmark$}           & \multicolumn{1}{c|}{}   & \multicolumn{1}{c|}{}   & \multicolumn{1}{c|}{}             &                      \\ \hline
				
				\citep{DBLP:conf/cikm/ZhouWHZMD22}& \multicolumn{1}{c|}{}                    & \multicolumn{1}{c|}{}   & \multicolumn{1}{c|}{}   & \multicolumn{1}{c|}{$\checkmark$}   & \multicolumn{1}{c|}{$\checkmark$}   & \multicolumn{1}{c|}{}   & \multicolumn{1}{c|}{}       & \multicolumn{1}{c|}{}                                     & \multicolumn{1}{c|}{$\checkmark$}      & \multicolumn{1}{c|}{}      &       & \multicolumn{1}{c|}{$\checkmark$}                     & \multicolumn{1}{c|}{}                          & \multicolumn{1}{c|}{}                            & \multicolumn{1}{c|}{}                        & \multicolumn{1}{c|}{}                     & \multicolumn{1}{c|}{}                      & \multicolumn{1}{c|}{}               & \multicolumn{1}{c|}{}   & \multicolumn{1}{c|}{$\checkmark$}   & \multicolumn{1}{c|}{}         &                      \\ \hline
				
				\citep{liu2020decoupled}& \multicolumn{1}{c|}{}                    & \multicolumn{1}{c|}{}   & \multicolumn{1}{c|}{}   & \multicolumn{1}{c|}{$\checkmark$}   & \multicolumn{1}{c|}{$\checkmark$}   & \multicolumn{1}{c|}{}   & \multicolumn{1}{c|}{}       & \multicolumn{1}{c|}{}                                     & \multicolumn{1}{c|}{$\checkmark$}      & \multicolumn{1}{c|}{}      &       & \multicolumn{1}{c|}{$\checkmark$}                     & \multicolumn{1}{c|}{}                          & \multicolumn{1}{c|}{}                            & \multicolumn{1}{c|}{}                        & \multicolumn{1}{c|}{}                     & \multicolumn{1}{c|}{}                      & \multicolumn{1}{c|}{}                & \multicolumn{1}{c|}{$\checkmark$}   & \multicolumn{1}{c|}{}   & \multicolumn{1}{c|}{}        &                      \\ \hline
				
				\citep{liu2021item}& \multicolumn{1}{c|}{}                    & \multicolumn{1}{c|}{}   & \multicolumn{1}{c|}{}   & \multicolumn{1}{c|}{}   & \multicolumn{1}{c|}{}   & \multicolumn{1}{c|}{$\checkmark$}   & \multicolumn{1}{c|}{}       & \multicolumn{1}{c|}{}                                     & \multicolumn{1}{c|}{$\checkmark$}      & \multicolumn{1}{c|}{}      &       & \multicolumn{1}{c|}{}                     & \multicolumn{1}{c|}{}                          & \multicolumn{1}{c|}{}                            & \multicolumn{1}{c|}{}                        & \multicolumn{1}{c|}{}                     & \multicolumn{1}{c|}{}                      & \multicolumn{1}{c|}{}              & \multicolumn{1}{c|}{$\checkmark$}   & \multicolumn{1}{c|}{}   & \multicolumn{1}{c|}{}          &      $\checkmark$                \\ \hline
				
				\citep{hao2020p}    & \multicolumn{1}{c|}{}     &\multicolumn{1}{c|}{$\checkmark$} &\multicolumn{1}{c|}{}  &\multicolumn{1}{c|}{$\checkmark$} &\multicolumn{1}{c|}{}   & \multicolumn{1}{c|}{}   & \multicolumn{1}{c|}{}       & \multicolumn{1}{c|}{}    & \multicolumn{1}{c|}{$\checkmark$}      & \multicolumn{1}{c|}{}      &      & \multicolumn{1}{c|}{$\checkmark$} & \multicolumn{1}{c|}{} &\multicolumn{1}{c|}{} &\multicolumn{1}{c|}{} &\multicolumn{1}{c|}{}                     &\multicolumn{1}{c|}{} & \multicolumn{1}{c|}{} & \multicolumn{1}{c|}{}   & \multicolumn{1}{c|}{$\checkmark$}   & \multicolumn{1}{c|}{}          &                     \\ \hline
				
				\citep{rakesh2019linked}    & \multicolumn{1}{c|}{}                         
				&\multicolumn{1}{c|}{} &\multicolumn{1}{c|}{}  &\multicolumn{1}{c|}{} &\multicolumn{1}{c|}{}   & \multicolumn{1}{c|}{$\checkmark$}   
				& \multicolumn{1}{c|}{}       & \multicolumn{1}{c|}{}                                     
				& \multicolumn{1}{c|}{}      & \multicolumn{1}{c|}{}      &      
				& \multicolumn{1}{c|}{} & \multicolumn{1}{c|}{} &\multicolumn{1}{c|}{} &\multicolumn{1}{c|}{} &\multicolumn{1}{c|}{}                     
				&\multicolumn{1}{c|}{} & \multicolumn{1}{c|}{} & \multicolumn{1}{c|}{}   & \multicolumn{1}{c|}{}   & \multicolumn{1}{c|}{}          &         $\checkmark$        \\ \hline
				\citep{wang2018path}& \multicolumn{1}{c|}{}                         
				&\multicolumn{1}{c|}{} &\multicolumn{1}{c|}{}  &\multicolumn{1}{c|}{} &\multicolumn{1}{c|}{}   & \multicolumn{1}{c|}{}   
				& \multicolumn{1}{c|}{$\checkmark$}       & \multicolumn{1}{c|}{}                                     
				& \multicolumn{1}{c|}{}      & \multicolumn{1}{c|}{}      &      
				& \multicolumn{1}{c|}{} & \multicolumn{1}{c|}{} &\multicolumn{1}{c|}{} &\multicolumn{1}{c|}{} &\multicolumn{1}{c|}{}                     
				&\multicolumn{1}{c|}{} & \multicolumn{1}{c|}{} & \multicolumn{1}{c|}{}   & \multicolumn{1}{c|}{}   & \multicolumn{1}{c|}{}          &    $\checkmark$                \\ \hline
				
				\citep{zhang2019identifying}  & \multicolumn{1}{c|}{}                         
				&\multicolumn{1}{c|}{} &\multicolumn{1}{c|}{$\checkmark$}  &\multicolumn{1}{c|}{} &\multicolumn{1}{c|}{}   & \multicolumn{1}{c|}{$\checkmark$}   
				& \multicolumn{1}{c|}{}       & \multicolumn{1}{c|}{}                                     
				& \multicolumn{1}{c|}{}      & \multicolumn{1}{c|}{}      &      
				& \multicolumn{1}{c|}{} & \multicolumn{1}{c|}{} &\multicolumn{1}{c|}{} &\multicolumn{1}{c|}{} &\multicolumn{1}{c|}{}                     
				&\multicolumn{1}{c|}{} & \multicolumn{1}{c|}{$\checkmark$} & \multicolumn{1}{c|}{}   & \multicolumn{1}{c|}{}   & \multicolumn{1}{c|}{}          &                     \\ \hline
				
				\citep{zhang2018quality}   & \multicolumn{1}{c|}{$\checkmark$}                         
				&\multicolumn{1}{c|}{$\checkmark$} &\multicolumn{1}{c|}{}  &\multicolumn{1}{c|}{} &\multicolumn{1}{c|}{}   & \multicolumn{1}{c|}{$\checkmark$}   
				& \multicolumn{1}{c|}{}       & \multicolumn{1}{c|}{}                                     
				& \multicolumn{1}{c|}{}      & \multicolumn{1}{c|}{}      &      
				& \multicolumn{1}{c|}{} & \multicolumn{1}{c|}{} &\multicolumn{1}{c|}{} &\multicolumn{1}{c|}{} &\multicolumn{1}{c|}{$\checkmark$}                     
				&\multicolumn{1}{c|}{} & \multicolumn{1}{c|}{} & \multicolumn{1}{c|}{}   & \multicolumn{1}{c|}{}   & \multicolumn{1}{c|}{}          &    $\checkmark$                 \\ \hline
				
				\citep{wan2018representing}  & \multicolumn{1}{c|}{}                         
				&\multicolumn{1}{c|}{} &\multicolumn{1}{c|}{}  &\multicolumn{1}{c|}{} &\multicolumn{1}{c|}{}   & \multicolumn{1}{c|}{}   
				& \multicolumn{1}{c|}{}       & \multicolumn{1}{c|}{$\checkmark$}                                     
				& \multicolumn{1}{c|}{}      & \multicolumn{1}{c|}{}      &      
				& \multicolumn{1}{c|}{} & \multicolumn{1}{c|}{} &\multicolumn{1}{c|}{} &\multicolumn{1}{c|}{} &\multicolumn{1}{c|}{}                     
				&\multicolumn{1}{c|}{} & \multicolumn{1}{c|}{} & \multicolumn{1}{c|}{$\checkmark$}   & \multicolumn{1}{c|}{}   & \multicolumn{1}{c|}{}          &                     \\ \hline
				
				\citep{kang2019complete}  & \multicolumn{1}{c|}{$\checkmark$}                         
				&\multicolumn{1}{c|}{} &\multicolumn{1}{c|}{}  &\multicolumn{1}{c|}{} &\multicolumn{1}{c|}{}   & \multicolumn{1}{c|}{}   
				& \multicolumn{1}{c|}{}       & \multicolumn{1}{c|}{}                                     
				& \multicolumn{1}{c|}{}      & \multicolumn{1}{c|}{}      &      
				& \multicolumn{1}{c|}{} & \multicolumn{1}{c|}{} &\multicolumn{1}{c|}{} &\multicolumn{1}{c|}{$\checkmark$} &\multicolumn{1}{c|}{}                     
				&\multicolumn{1}{c|}{} & \multicolumn{1}{c|}{} & \multicolumn{1}{c|}{}   & \multicolumn{1}{c|}{}   & \multicolumn{1}{c|}{}          &                     \\ \hline
				
				\citep{entezari2021tensor}  & \multicolumn{1}{c|}{}                         
				&\multicolumn{1}{c|}{} &\multicolumn{1}{c|}{}  &\multicolumn{1}{c|}{} &\multicolumn{1}{c|}{}   & \multicolumn{1}{c|}{}   
				& \multicolumn{1}{c|}{}       & \multicolumn{1}{c|}{$\checkmark$}                                     
				& \multicolumn{1}{c|}{}      & \multicolumn{1}{c|}{}      &      
				& \multicolumn{1}{c|}{} & \multicolumn{1}{c|}{} &\multicolumn{1}{c|}{} &\multicolumn{1}{c|}{} &\multicolumn{1}{c|}{}                     
				&\multicolumn{1}{c|}{} & \multicolumn{1}{c|}{} & \multicolumn{1}{c|}{}   & \multicolumn{1}{c|}{}   & \multicolumn{1}{c|}{}          & $\checkmark$                   \\ \hline
				
				\citep{ma2021neat}  & \multicolumn{1}{c|}{}                         
				&\multicolumn{1}{c|}{} &\multicolumn{1}{c|}{}  &\multicolumn{1}{c|}{} &\multicolumn{1}{c|}{}   & \multicolumn{1}{c|}{}   
				& \multicolumn{1}{c|}{}       & \multicolumn{1}{c|}{$\checkmark$}                                     
				& \multicolumn{1}{c|}{}      & \multicolumn{1}{c|}{}      &      
				& \multicolumn{1}{c|}{} & \multicolumn{1}{c|}{} &\multicolumn{1}{c|}{} &\multicolumn{1}{c|}{} &\multicolumn{1}{c|}{}                     
				&\multicolumn{1}{c|}{} & \multicolumn{1}{c|}{} & \multicolumn{1}{c|}{}   & \multicolumn{1}{c|}{}   & \multicolumn{1}{c|}{}          &    $\checkmark$                \\ \hline
				
				\citep{angelovska2021siamese}  & \multicolumn{1}{c|}{}                         
				&\multicolumn{1}{c|}{$\checkmark$} &\multicolumn{1}{c|}{}  &\multicolumn{1}{c|}{} &\multicolumn{1}{c|}{}   & \multicolumn{1}{c|}{}   
				& \multicolumn{1}{c|}{}       & \multicolumn{1}{c|}{}                                     
				& \multicolumn{1}{c|}{}      & \multicolumn{1}{c|}{}      &      
				& \multicolumn{1}{c|}{} & \multicolumn{1}{c|}{} &\multicolumn{1}{c|}{} &\multicolumn{1}{c|}{} &\multicolumn{1}{c|}{$\checkmark$}                     
				&\multicolumn{1}{c|}{$\checkmark$} & \multicolumn{1}{c|}{} & \multicolumn{1}{c|}{}   & \multicolumn{1}{c|}{}   & \multicolumn{1}{c|}{}          &                     \\ \hline
				
				\citep{ma2023personalized}  & \multicolumn{1}{c|}{}                         
				&\multicolumn{1}{c|}{} &\multicolumn{1}{c|}{}  &\multicolumn{1}{c|}{} &\multicolumn{1}{c|}{}   & \multicolumn{1}{c|}{}   
				& \multicolumn{1}{c|}{}       & \multicolumn{1}{c|}{$\checkmark$}                                     
				& \multicolumn{1}{c|}{}      & \multicolumn{1}{c|}{}      &      
				& \multicolumn{1}{c|}{} & \multicolumn{1}{c|}{} &\multicolumn{1}{c|}{$\checkmark$} &\multicolumn{1}{c|}{} &\multicolumn{1}{c|}{}                     
				&\multicolumn{1}{c|}{} & \multicolumn{1}{c|}{} & \multicolumn{1}{c|}{}   & \multicolumn{1}{c|}{}   & \multicolumn{1}{c|}{}          &                     \\ \hline
				
				\citep{yang2022inferring}  & \multicolumn{1}{c|}{}                         
				&\multicolumn{1}{c|}{$\checkmark$} &\multicolumn{1}{c|}{}  &\multicolumn{1}{c|}{$\checkmark$} &\multicolumn{1}{c|}{}   & \multicolumn{1}{c|}{$\checkmark$}   
				& \multicolumn{1}{c|}{}       & \multicolumn{1}{c|}{}                                     
				& \multicolumn{1}{c|}{}      & \multicolumn{1}{c|}{$\checkmark$}      &     
				& \multicolumn{1}{c|}{$\checkmark$} & \multicolumn{1}{c|}{} &\multicolumn{1}{c|}{} &\multicolumn{1}{c|}{} &\multicolumn{1}{c|}{}                     
				&\multicolumn{1}{c|}{} & \multicolumn{1}{c|}{} & \multicolumn{1}{c|}{}   & \multicolumn{1}{c|}{}   & \multicolumn{1}{c|}{}          &                     \\ \hline
				
				\citep{wu2019session}  & \multicolumn{1}{c|}{$\checkmark$}                         
				&\multicolumn{1}{c|}{} &\multicolumn{1}{c|}{}  &\multicolumn{1}{c|}{} &\multicolumn{1}{c|}{}   & \multicolumn{1}{c|}{$\checkmark$}   
				& \multicolumn{1}{c|}{$\checkmark$}       & \multicolumn{1}{c|}{}                                     
				& \multicolumn{1}{c|}{}      & \multicolumn{1}{c|}{}      &     
				& \multicolumn{1}{c|}{} & \multicolumn{1}{c|}{} &\multicolumn{1}{c|}{} &\multicolumn{1}{c|}{} &\multicolumn{1}{c|}{}                     
				&\multicolumn{1}{c|}{} & \multicolumn{1}{c|}{} & \multicolumn{1}{c|}{}   & \multicolumn{1}{c|}{}   & \multicolumn{1}{c|}{}          &    $\checkmark$               \\ \hline
				\citep{xu2020product}  & \multicolumn{1}{c|}{}                      
				&\multicolumn{1}{c|}{} &\multicolumn{1}{c|}{}  &\multicolumn{1}{c|}{$\checkmark$} &\multicolumn{1}{c|}{}   & \multicolumn{1}{c|}{$\checkmark$}   
				& \multicolumn{1}{c|}{}       & \multicolumn{1}{c|}{}                                     
				& \multicolumn{1}{c|}{}      & \multicolumn{1}{c|}{$\checkmark$}      &     
				& \multicolumn{1}{c|}{$\checkmark$} & \multicolumn{1}{c|}{} &\multicolumn{1}{c|}{} &\multicolumn{1}{c|}{} &\multicolumn{1}{c|}{}                     
				&\multicolumn{1}{c|}{} & \multicolumn{1}{c|}{} & \multicolumn{1}{c|}{}   & \multicolumn{1}{c|}{}   & \multicolumn{1}{c|}{}          &    $\checkmark$            \\ \hline
				\citep{zheng2009substitutes}  & \multicolumn{1}{c|}{}                      
				&\multicolumn{1}{c|}{} &\multicolumn{1}{c|}{}  &\multicolumn{1}{c|}{} &\multicolumn{1}{c|}{}   & \multicolumn{1}{c|}{$\checkmark$}   
				& \multicolumn{1}{c|}{$\checkmark$}       & \multicolumn{1}{c|}{}                                     
				& \multicolumn{1}{c|}{}      & \multicolumn{1}{c|}{}      &     
				& \multicolumn{1}{c|}{} & \multicolumn{1}{c|}{} &\multicolumn{1}{c|}{} &\multicolumn{1}{c|}{} &\multicolumn{1}{c|}{}                     
				&\multicolumn{1}{c|}{} & \multicolumn{1}{c|}{} & \multicolumn{1}{c|}{}   & \multicolumn{1}{c|}{}   & \multicolumn{1}{c|}{}          &    $\checkmark$                 \\ \hline
				\citep{DBLP:journals/corr/abs-2401-02130}  & \multicolumn{1}{c|}{}                      
				&\multicolumn{1}{c|}{} &\multicolumn{1}{c|}{}  &\multicolumn{1}{c|}{$\checkmark$} &\multicolumn{1}{c|}{}   & \multicolumn{1}{c|}{$\checkmark$}   
				& \multicolumn{1}{c|}{}       & \multicolumn{1}{c|}{}                                     
				& \multicolumn{1}{c|}{}      & \multicolumn{1}{c|}{}      &  $\checkmark$   
				& \multicolumn{1}{c|}{} & \multicolumn{1}{c|}{} &\multicolumn{1}{c|}{} &\multicolumn{1}{c|}{} &\multicolumn{1}{c|}{}                     
				&\multicolumn{1}{c|}{} & \multicolumn{1}{c|}{} & \multicolumn{1}{c|}{$\checkmark$}   & \multicolumn{1}{c|}{}   & \multicolumn{1}{c|}{}          &     $\checkmark$                \\ \hline
				\citep{DBLP:journals/www/LiCH21}  & \multicolumn{1}{c|}{$\checkmark$}                      
				&\multicolumn{1}{c|}{} &\multicolumn{1}{c|}{}  &\multicolumn{1}{c|}{$\checkmark$} &\multicolumn{1}{c|}{}   & \multicolumn{1}{c|}{}   
				& \multicolumn{1}{c|}{}       & \multicolumn{1}{c|}{}                                     
				& \multicolumn{1}{c|}{}      & \multicolumn{1}{c|}{}      &   
				& \multicolumn{1}{c|}{} & \multicolumn{1}{c|}{} &\multicolumn{1}{c|}{} &\multicolumn{1}{c|}{} &\multicolumn{1}{c|}{$\checkmark$  }                     
				&\multicolumn{1}{c|}{} & \multicolumn{1}{c|}{} & \multicolumn{1}{c|}{}   & \multicolumn{1}{c|}{}   & \multicolumn{1}{c|}{}          &                \\ \hline
				\citep{DBLP:conf/ijcnn/TkachukWDL22}  & \multicolumn{1}{c|}{}                      
				&\multicolumn{1}{c|}{$\checkmark$} &\multicolumn{1}{c|}{}  &\multicolumn{1}{c|}{} &\multicolumn{1}{c|}{$\checkmark$}   & \multicolumn{1}{c|}{$\checkmark$}   
				& \multicolumn{1}{c|}{}       & \multicolumn{1}{c|}{}                                     
				& \multicolumn{1}{c|}{}      & \multicolumn{1}{c|}{}      &  $\checkmark$  
				& \multicolumn{1}{c|}{} & \multicolumn{1}{c|}{} &\multicolumn{1}{c|}{} &\multicolumn{1}{c|}{} &\multicolumn{1}{c|}{}                     
				&\multicolumn{1}{c|}{} & \multicolumn{1}{c|}{} & \multicolumn{1}{c|}{}   & \multicolumn{1}{c|}{}   & \multicolumn{1}{c|}{}          &        $\checkmark$           \\ \hline
			\end{tabular}%
		}
	\end{table}
\end{landscape}
The formulas (\ref{eqn:301}) and (\ref{eqn:302}) are used to capture complementary relationships, and the formulas (\ref{eqn:30}) and (\ref{eqn:303}) are used to capture substitution relationships. \citep{rakesh2019linked}proposed a Link Variational Autoencoder(LVA), which uses a link predictor to learn the relationship between products; \citep{zhao2017deep}uses the sigmoid function to calculate the complementary probability, such as the formula (\ref{eqn:5}) shown.
\begin{equation}
	\begin{split}
		P(y=1|q,c)&=\sigma(x^{T}_{q}Mx_{c}+b)\\
		&=\frac{1}{1+e^{-(x^{T}_{q}Mx_{c}+b)}} \label{eqn:5}
	\end{split}
\end{equation}
where M is the compatibility matrix and b is the bias term, which is continuously optimized during training. The $x^{T}_{q}M$ represents the most compatible product with product q.
\newline \indent \textbf{Learning complementary relationships based on image content. }Use product images to learn the compatibility of two products.
In \citep{DBLP:journals/www/LiCH21}, ResNet-18 is used to extract high-level visual features from the original input images. The CNN module is then fine-tuned through the item classification task. The principle behind this approach is that different fashion items often exhibit distinct attribute distributions. For example, "sleeve length" is an important attribute for shirts and sweaters, while people tend to pay more attention to the "waist circumference" of dresses. This requires the model to focus on different attributes when dealing with different types of clothing.	 \citep{kang2019complete}uses ResNet-50 to extract features from product image $I_{p}$ and scene image $I_{s}$, and obtain embedding vectors $v_{p}$ and $v_{s}$. Use two layers of FNN to convert the embedding vector into a unified style space to obtain the style embedding of products and scenes. As follows:
\begin{equation}
	\mathbf{f}_{s}=g\left(\Theta_{g} ; \mathbf{v}_{s}\right), \mathbf{f}_{p}=g\left(\Theta_{g} ; \mathbf{v}_{p}\right)\label{eqn:68}
\end{equation}
Use the distance function to measure the global compatibility of products and scenes:
\begin{equation}
	d_{\text {global }}(s, p)=\left\|\mathbf{f}_{s}-\mathbf{f}_{p}\right\|^{2}\label{eqn:69}
\end{equation}
According to the feature map $m_{i}$, the embedding of the $i$-th region in the scene image is obtained:
\begin{equation}
	\mathbf{f}_{i}=g\left(\Theta_{l} ; \mathbf{m}_{i}\right), \hat{\mathbf{f}}_{i}=g\left(\Theta_{\hat{l}} ; \mathbf{m}_{i}\right)\label{eqn:70}
\end{equation}
Scene images are usually large regions that include many contents, and considering only global compatibility may overlook key details in the scene. Therefore, each region of the scene images is matched with the product images and Attention is used to assign different weights to each region for a more fine-grained matching process.
\begin{equation}
	d_{\text {local }}(s, p)=\sum_{1 \leq i \leq w \times h} a_{i}\left\|\mathbf{f}_{i}-\mathbf{f}_{p}\right\|^{2}\label{eqn:70}
\end{equation}
Finally, a mixing distance is defined to measure the compatibility of products and scenes:
\begin{equation}
	d_{*}(s, p)=\frac{1}{2}\left[d_{\text {global }}(s, p)+d_{\text {local }}(s, p)\right]\label{eqn:70}
\end{equation}
\newline \indent \textbf{Combining text content and image content to learn complementary relationships. }For different categories of products, the complementarity situation is also different. For example, for clothing, matching is mainly based on style, so the style needs to be judged based on pictures of clothes; for computers and chargers, function matching needs to be based on interfaces, so product description information is needed. Based on this, \citep{DBLP:conf/www/BibasSJ23,zhang2018quality,DBLP:conf/recsys/Papso23}combines the two, using both the text content of the product and the image of the product. For feature extraction of product images, \citep{zhang2018quality}uses CNN, and the selected feature is the output of the second fully connected layer in the CNN. \citep{DBLP:conf/www/BibasSJ23}uses the pre-trained model ResNet152 and uses Multilayer Perceptron(MLP) for output. For feature extraction of product text information, \citep{zhang2018quality}uses distributed representation to obtain text embedding vectors based on the title and description of the product; \citep{DBLP:conf/www/BibasSJ23}for categorical feature(discrete feature), each value has a corresponding embedding, for continuous features, discretize them into categorical features. \citep{DBLP:conf/recsys/Papso23}directly uses the image and text information metadata of the product to fine-tune the contrastive learning to obtain a universal embedding space for the product. \citep{wu2019session}fuses the basic embedding $m_{i}$ of the product, the category embedding $g_{i}$ and the event embedding $t_{i}$ together to obtain the final embedding, as shown below :
\begin{equation}
	d_{i}=Concat([m_{i},m_{i}\odot g_{i},g_{i}])\label{eqn:299}
\end{equation}
\begin{equation}
	x_{i}=elu(W^{T}_{1} elu(W^{T}_{2}d_{i}+b_{2})+b_{1})\odot (1+t_{i})\label{eqn:300}
\end{equation}
The basic embedding $m_{i}$ can be randomly initialized or represented by the image embedding of the product. Using the nonlinear activation function Relu to fuse category features into product embeddings not only allows the model to better expand to new or rare products, but also improves prediction accuracy by better capturing the relationship between different categories.
\newline \indent In terms of relationship prediction, \citep{zhang2018quality}calculates the Euclidean distance between the image feature vectors $m_{i}$, $m_{j}$ of two products $i$, $j$ in the embedding space. The formula is:
\begin{equation}
	d_{j \mid i}^{(c m)}\left(I_{i}, I_{j}\right)=\left\|\left(\mathbf{m}_{i}-\mathbf{m}_{j}\right)^{T} \mathbf{E}_{M}\right\|_{2}^{2}\label{eqn:71}
\end{equation}
Where $\mathbf{E}_{M}$ is the low-order Mahalanobis transformation matrix. And calculate the Euclidean distance of the text embedding vector:
\begin{equation}
	d_{j \mid i}^{(c t)}\left(I_{i}, I_{j}\right)=\left\|\left(\mathbf{t}_{i}-\mathbf{t}_{j}\right)^{T} \mathbf{E}_{T}\right\|_{2}^{2}\label{eqn:72}
\end{equation}
Where $\mathbf{E}_{T}$ is the training vector used to learn the text features of product complementary relationships, and the final product complementary distance is obtained by combining the two types of distances. 
\newline \indent \citep{DBLP:conf/www/BibasSJ23}connects all features and outputs them through MLP. \citep{DBLP:conf/www/BibasSJ23}not only learns the complementary relationship between products, but also learns the complementary relationship between categories. Use MLP to project product embeddings into specific categories, and then use triplet loss for optimization; \citep{DBLP:conf/recsys/Papso23}in the product universal embedding space, two similar products should be closer to each other. Then combine it with user purchase data to explore product complementary relationships.
\subsubsection{Learning complementary relationships based on user purchase sequences}
\begin{figure}[htb]
	\centering
	\includegraphics[scale=0.5]{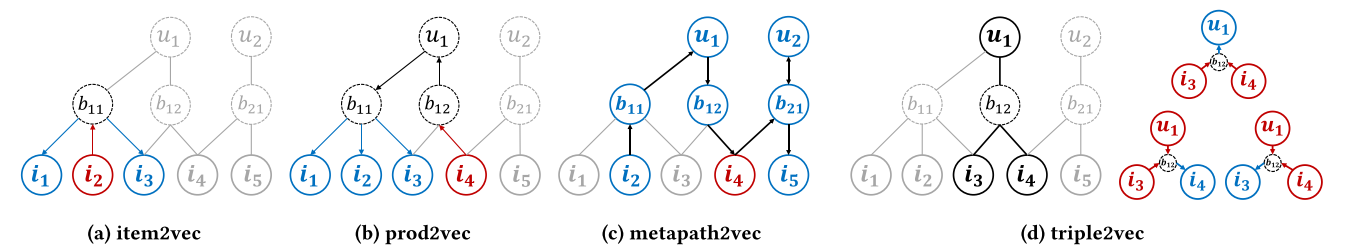}
	\caption{Comparison of embedding methods}\label{fig:fig4}
\end{figure}

\
\newline Learn product embeddings based on users' purchase sequences for relationship prediction.
\newline \indent \textbf{Learn complementary relationships based on the entire purchase sequence.} In terms of feature extraction, \citep{wang2018path}uses Item2vec. Item2vec draws on word2vec, treating products as words to learn the relationship between products. The Item2vec method focuses on the co-purchase relationship of products in the purchase sequence, without considering the user. \citep{trofimov2018inferring,kvernadze2022two}uses Prod2vec, which emphasizes the relationship between each user's cross-sequence items. As shown in Figure\ref{fig:fig4}, the similarities and differences between Item2vec and Prod2vec are shown.
\newline \indent In terms of predicting the complementary relationship between products, \citep{wang2018path}uses mapping function to map the initial embedding vector of products to substitutable space and complementary space respectively, and proposes a path constraint method that combines product embeddings with path constraints to further improve the model performance. The first is the category constraint. If two products are related, then the categories to which the two products belong are also related.
\begin{equation}
	(Prod_{A},RelatedTo,Prod_{B})\Rightarrow (Category_{A},RelatedTo,Category_{B})\label{eqn:6}
\end{equation}
\indent Next is the two-step path constraint, in the form $\forall i,j,k:(i,r_{1},j)\wedge (j,r_{2},k)\Rightarrow (i,r_{ 3},k)$. It is usually represented in the product relationship graph in the following form:
\begin{equation}
	\begin{split}
		(Prod_{A},Subst,Prod_{B})& \wedge (Prod_{B},Subst,Prod_{C})\\
		& \Rightarrow (Prod_{A},Subst,Prod_{C}) \label{eqn:7}
	\end{split}
\end{equation}
\begin{equation}
	\begin{split}
		(Prod_{A},Compl,Prod_{B})& \wedge (Prod_{B},Subst,Prod_{C})\\
		& \Rightarrow (Prod_{A},Compl,Prod_{C}) \label{eqn:8}
	\end{split}
\end{equation}
\begin{equation}
	\begin{split}
		(Prod_{A},Subst,Prod_{B})& \wedge (Prod_{B},Compl,Prod_{C})\\
		& \Rightarrow (Prod_{A},Compl,Prod_{C}) \label{eqn:9}
	\end{split}
\end{equation}
\begin{equation}
	\begin{split}
		(Prod_{A},Compl,Prod_{B})& \wedge (Prod_{B},Compl,Prod_{C})\\
		& \Rightarrow (Prod_{A},Compl,Prod_{C}) \label{eqn:10}
	\end{split}
\end{equation}
However this method may ignore useful dependent products in longer paths that can help infer product relationships.
\newline \indent \textbf{Learning complementary relationships based on user-purchased product triplets.} \citep{wan2018representing,ma2023personalized,entezari2021tensor}models the triplet formed by two items purchased by the user in the same shopping basket. \citep{wan2018representing,ma2023personalized}mainly uses triple2vec to learn product feature vectors and user feature vectors simultaneously. As shown in Figure\ref{fig:fig4}, unlike existing product representations based on word2vec, triple2vec focuses on the close relationship of (item, item, user), and the triple is composed of the following form:
\begin{equation}
	\mathcal{T}=\left\{(i, j, u) \mid i \in \mathcal{I}_{b} \wedge j \in \mathcal{I}_{b} \wedge i \neq j \wedge b \in \mathcal{B}_{u} \wedge u \in \mathcal{U}\right\}\label{eqn:81}
\end{equation}
Define the cohesion score of each triple $(i,j,u)$ as:
\begin{equation}
	s_{i, j, u}=f_{i}^{T}{g}_{j}+f_{i}^{T}{h}_{u}+{g}_{j}^{T}{h}_{u}\label{eqn:82}
\end{equation}
Triple2vec learns two representation sets $f$ and $g$ of the product. These two embeddings describe the functions and attributes of the product from different perspectives. But the inner product between the two captures the complementarity between products. Among them, $f_{i}^{T}{g}_{j}$ is used to model the complementarity between two products in the same shopping basket. $f_{i}^{T}{h}_{u}+{g}_{j}^{T}{h}_{u}$ is used to capture the compatibility between the product and the user, that is, the product the degree to which potential features match user preferences. A higher cohesion score indicates a closer connection between the nodes in the triplet.
\newline \indent \citep{entezari2021tensor} models the product triplet purchased by the user as a three-mode tensor $\underline{X}$. Each element of the three-mode tensor represents the number of times the user purchased two products together. By decomposing this tensor, product embeddings and user embeddings can be learned, as shown below:
\begin{equation}
	\underline{X}\approx A \underline{R}A^{T}\label{eqn:83}
\end{equation}
Among them, A is the product embedding matrix, and R is the user embedding matrix. The product embedding matrix A is shared among all users. By taking the dot product of this matrix and its transpose, the products most frequently purchased together will receive a higher score.
\newline \indent \citep{ma2021neat}extracts the Gaussian embedding of the product, and use the average vector of each product as a feature representation.
\newline \indent In terms of predicting the complementary relationship between products, \citep{ma2021neat}uses the cosine similarity between the average vectors of two products to represent;
\subsubsection{Learn complementary relationships based on product relationship graphs}
\begin{figure}[htb]
	\centering
	\includegraphics[scale=0.8]{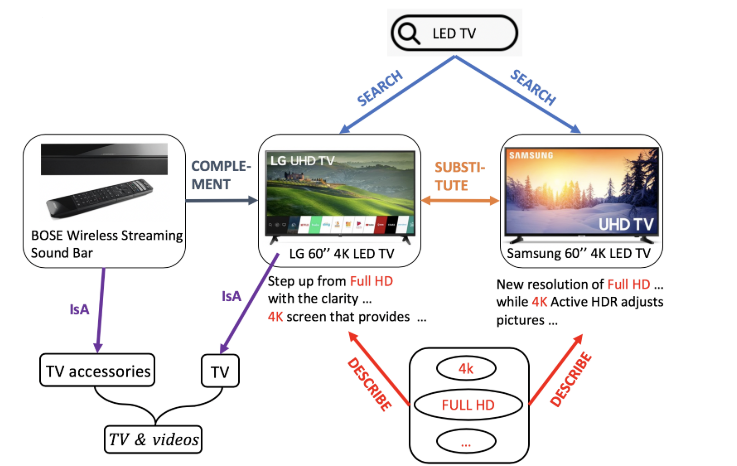}
	\caption{Product Knowledge Graph}\label{fig:fig81}
\end{figure}

\
\newline It may not be possible to learn an accurate representation of product relationships by relying solely on product content, so a product complementary relationship graph and a substitution relationship graph are combined based on the product content. Convert reasoning about the relationship between products into a link prediction task.
\newline \indent \textbf{Graph type.} The papers \citep{liu2020decoupled,DBLP:conf/aaai/ChenHXFLSYQ23,DBLP:conf/cikm/ZhouWHZMD22,zhang2021learning,hao2020p,liu2021item,yang2022inferring} employ a product relationship heterogeneous graph. Among them, \citep{liu2020decoupled,DBLP:conf/aaai/ChenHXFLSYQ23,DBLP:conf/cikm/ZhouWHZMD22,zhang2021learning} are heterogeneous undirected graphs, which are decoupled into complementary relationship undirected graphs and substitute relationship undirected graphs. \citep{hao2020p,liu2021item,yang2022inferring} use a heterogeneous directed graph. \citep{yan2022personalized,DBLP:journals/corr/abs-2401-02130,DBLP:conf/ijcnn/TkachukWDL22} adopt a product complementary relationship graph, where \citep{yan2022personalized} uses a commodity complementary directed graph,  \citep{DBLP:journals/corr/abs-2401-02130} uses a commodity complementary undirected graph,and \citep{DBLP:conf/ijcnn/TkachukWDL22} uses a hypergraphs. \citep{yang2022inferring,xu2020product} utilize a knowledge graph containing commodity relationships and attributes, structured as shown in Figure \ref{fig:fig81}.
\newline \indent  \textbf{Initial Embedding.} \citep{liu2020decoupled,DBLP:conf/aaai/ChenHXFLSYQ23,DBLP:conf/cikm/ZhouWHZMD22,yang2022inferring}uses a MLP for initial embedding of textual attributes such as category, brand, etc. of the product. \citep{hao2020p}use 3-layer FNN for initial embedding of the product's ID, category, and title; \citep{zhang2021learning}uses the user's purchase sequence of the product, and uses the embedding matrix to transform the one-hot encoding of the product to obtain the initial embedding; \citep{yan2022personalized}uses a generic embedding module consisting of two FNNs + one BN to get the initial embedding of the product.\citep{DBLP:journals/corr/abs-2401-02130} uses BERT as the pretraining model to obtain category embeddings and discretize the continuous price into bins using equal-depth binning.
\newline \indent \textbf{Use Graph Neural Network for relationship learning.} \citep{liu2020decoupled}uses GraphSage; \citep{DBLP:conf/aaai/ChenHXFLSYQ23}uses PinSage; \citep{DBLP:conf/cikm/ZhouWHZMD22,yan2022personalized}uses GAT; \citep{hao2020p}uses Product2vec, which is a one-hop GAT; \citep{liu2021item}uses multi-hop GNN pairs; \citep{zhang2021learning}uses interactive GCN. \citep{DBLP:journals/corr/abs-2401-02130}uses spectral-
based GCN filters. \citep{DBLP:conf/ijcnn/TkachukWDL22} utilizes a novel hypergraph embedding method called Cleora. Hypergraph methods outperform traditional graph approaches significantly in product embedding, website embedding, and similar data domains. They exhibit greater expressive power and better capture the underlying data generation processes.
\newline \indent \textbf{Knowledge Graph reasoning.} \citep{yang2022inferring}uses Markov Decision Process (MDP) to reason about the relationship of products in the knowledge graph, fine-grained reasoning can be performed, and the reasoning is interpretable. \citep{xu2020product}proposes a representation learning model based on self-attention, which is used to automatically retrieve relationships in product knowledge graphs and learn product embeddings from user activity data and product information in an end-to-end manner.
\subsection{Complex complementary relationships}
\subsubsection{Complementary relationship is asymmetrical}
\
\newline For the complementary relationship between products, there may be asymmetry in the complementary relationship. When a user buys a cellphone, you can recommend him a charger and a tempered film as a complementary product to the cellphone; but when a user buys a charger, it makes no sense to recommend him a cellphone. \citep{mcauley2015inferring,wu2022towards,wang2018path,hao2020p}proposed a solution to this problem.
\newline \indent \citep{mcauley2015inferring}first predict whether there is a complementary relationship between two products, and then predict the direction of the complementary relationship. It can be written in the following form:
\begin{equation}
	p((a, b) \in \mathcal{E})=\underbrace{p(a \leftrightarrow b)}_{\text {'are they related?' }}\overbrace{p(a \rightarrow b \mid a \leftrightarrow b)}^{\text {'does the edge flow in this direction?' }}\label{eqn:101}
\end{equation}
The complementary relationship between products is based on the inner product of the two product embedding vectors $\psi_{\theta}(i, j)=\left(1, \theta_{i, 1} \cdot \theta_{j, 1} , \theta_{i, 2} \cdot \theta_{j, 2}, \ldots, \theta_{i, K} \cdot \theta_{j, K}\right)$ to calculate. The direction of the complementary relationship is based on the difference between the two product embedding vectors $\varphi_{\theta}(i, j)=\left(1, \theta_{j, 1}-\theta_{i, 1}, \ldots, \ theta_{j, K}-\theta_{i, K}\right)$ to calculate.
\newline \indent \citep{wang2018path}generates two feature vectors for item $i$: query vector $v_{i}$ and complementary vector $v_{i}^{\prime}$. The query vector is the feature vector used when the product is used as a query product, and the complementary vector is the feature vector used when the product is used as a complement of other products. First predict the direction of the relationship between two products, as shown in the formula (\ref{eqn:120}), and then predict the type of this relationship.
\begin{equation}
	P\left(y_{i, j} \mid V, V^{\prime}\right)=\sigma\left(v_{i}^{T} \cdot v_{j}^{\prime}\right)\label{eqn:120}
\end{equation}
The sigmoid activation function used.
\newline \indent\citep{hao2020p}generate query vectors and complementary vectors for each product category to solve the problem of asymmetric complementary relationships. \citep{wu2022towards}proposed a logical reasoning module that uses a combination of three basic logical operations (projection, intersection, negation) to automatically filter out product pairs with substitution relationships from co-purchase data. The defined logical operators (except negation) are directed and irreversible, thus accommodating the asymmetric nature of the recommendations.
\subsubsection{Higher order complementarity}
\
\newline Compared with querying the complementary products of a single product, the complementary products of the product combination query set should be more comprehensive and meaningful. For example, if a user purchases a computer, mouse, and keyboard, he may plan to build an office, so the probability of recommending a desk should be higher at this time. \citep{wu2022towards}uses Attention to learn high-order dependencies between the query product set and potentially complementary products. Specifically, the Attention automatically learns the weights of each product in the query product set and then aggregates them into a unique representation of the entire query set.
\subsubsection{Complementary substitution relationships influence each other}
\begin{figure}[htb]
	\centering
	\includegraphics[scale=0.8]{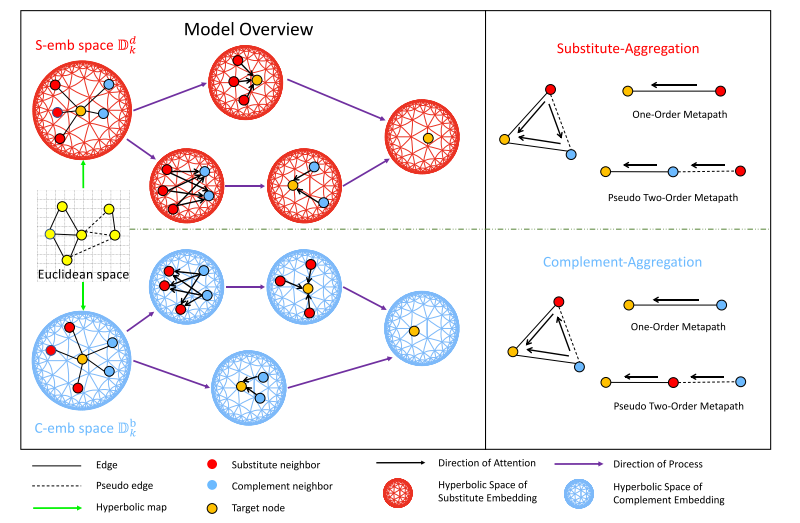}
	\caption{literature \citep{DBLP:conf/cikm/ZhouWHZMD22} Structure diagram}\label{fig:fig101}
\end{figure}

\begin{figure}[htb]
	\centering
	\includegraphics[scale=0.4]{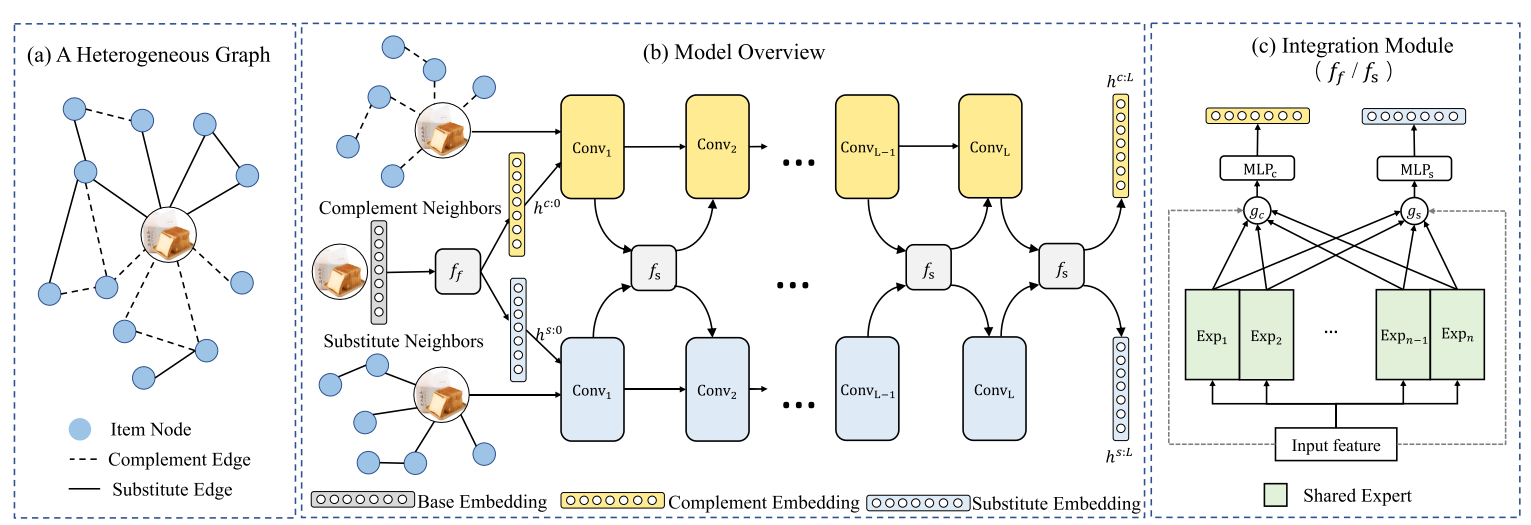}
	\caption{literature \citep{DBLP:conf/aaai/ChenHXFLSYQ23} Structure diagram}\label{fig:fig102}
\end{figure}

\
\newline Useful complementary information can also be obtained in the substitute information of products. For example, two substitutable products may have many identical complementary products. \citep{liu2020decoupled,DBLP:conf/aaai/ChenHXFLSYQ23,DBLP:conf/cikm/ZhouWHZMD22}focuses on the interaction between product complementary relationships product substitution relationships. Decouple the heterogeneous product relationship graph into an undirected graph of complementary relationships and an undirected graph of substitution relationships. Among them, \citep{liu2020decoupled,DBLP:conf/aaai/ChenHXFLSYQ23}are decoupled into Euclidean space. \citep{DBLP:conf/cikm/ZhouWHZMD22}is decoupled into hyperbolic space and is considered to have stronger expressive power than Euclidean space,then perform fusion learning on the two relationship graphs, as shown in Figure\ref{fig:fig101}.
\newline \indent \citep{liu2020decoupled,DBLP:conf/aaai/ChenHXFLSYQ23}proposed a two-step integration strategy. \citep{liu2020decoupled}proposed the DecGCN model, which uses the graph convolutional neural network GraphSage for graph learning. The first step of structural integration uses a joint attention neighborhood aggregation strategy in each convolutional layer to integrate substitute neighborhoods and complementary neighborhoods. As shown in the formulas(\ref{eqn:11})-(\ref{eqn:12}). For learning the complementary relationship of products, only the structural information of the product substitution graph is used, that is, the substitute neighborhoods of the product. But feature representation of the substitute neighborhoods of the product still uses complementary feature vectors, which means that the semantic information of the substitute neighbor is not used. The second step is semantic integration, which enables knowledge transfer between the semantics of two different relationships. Such as the formula(\ref{eqn:102}), the knowledge of substitute neighbors is transferred to complementary neighbors.
\begin{equation}
	\widetilde{Z}^{(s)}=GCN^{(s)}(G^{(s)}|G^{(c)},X)\label{eqn:11}
\end{equation}
\begin{equation}
	\widetilde{Z}^{(c)}=GCN^{(c)}(G^{(c)}|G^{(s)},X)\label{eqn:12}
\end{equation}
Among them, $G^{(s)}$ and $G^{(c)}$ are the product substitution relationship graph and the complementation relationship graph respectively.
\begin{equation}
	\hat{\mathbf{z}}_{i}^{\prime(c)}=\left(1-\alpha^{\prime}\right) \tilde{\mathbf{z}}_{i}^{(c)}+\alpha^{\prime} f_{s \rightarrow c}\left(\tilde{\mathbf{z}}_{i}^{(s)}\right)\label{eqn:102}
\end{equation}
Where $f_{s \rightarrow c}$ is a 3-layer MLP.
\newline \indent DecGCN works effectively when the product has both substitute and complementary neighbors. If a product has only one type of neighbors (substitute neighbors or complementary neighbors), which is common in real situations, then DeCGCN's integration scheme will be ineffective. \citep{DBLP:conf/aaai/ChenHXFLSYQ23}is improved on the basis of \citep{liu2020decoupled}. As shown in the figure \ref{fig:fig102}, the Graph Convolution Network PinSage and the multi-task learning model MMOE are combined, so that products can learn complementary relationships and substitution relationships at the same time. The first step is to use feature integration at the bottom to generate feature representations of different relationships through the MMOE model based on the initial features of the product. The two relationships of the product share the feature encoding function. It can work effectively even if a node has only one type of neighbor. The second step is structural integration, which aggregates and splices the product's substitute neighbors and complementary neighbors together to obtain neighbor information, and then obtains the integrated substitute neighbor information and complementary neighbor information through the MMOE model. Finally, we can get the complementary relationship representation of the product that incorporates the substitution relationship and the substitution relationship representation that incorporates the complementary relationship.
\newline \indent \citep{DBLP:conf/cikm/ZhouWHZMD22}uses hyperbolic Pseudo Two-Order Metapath aggregation to propagate the mutual influence between two different relationships. The metapath constructed in the complementary space is shown as formula (\ref{eqn:103}).
\begin{equation}
	\text { target node } \stackrel{s}{\longleftarrow} \text { s neighbors } \stackrel{c}{\leftarrow- -} \text { c neighbors }\label{eqn:103}
\end{equation}
Next, calculate the influence of each complementary neighbor on the substitute neighbor, aggregate the complementary neighbors into each substitute neighbor, and obtain the updated product complementary relationship embedding, and then use the product complementary relationship embedding to obtain the hyperbolic embedding of the product complementary relationship. As follows:
\begin{equation}
	\widetilde{z}_{m, t}^{(c)}=\sigma\left(\left(\alpha \odot z_{m}^{(c)}\right) \oplus_{k} \exp _{0}^{k}\left(\sum_{v_{i} \in \mathcal{N}_{m}^{(c)}} \log _{0}^{k}\left(e_{m i} \odot_{k} \widetilde{z}_{i}^{(c)}\right)\right)\right)\label{eqn:104}
\end{equation}
\subsubsection{There are multiple relationships between products}
\
\newline In real life, there may be both complementary and substitutive relationships between two products. For example, two cellphone cases can be viewd together and purchased after view(substitution relationship) or purchased together (complementary relationship), because they are substitutes in function and complementary in color and texture. \citep{wu2022towards,hao2020p}treat them as substitution relations and remove them from the set of complementary relations. \citep{liu2021item}use multiple relationships of product pairs at the same time to generate relationship vector $e_{w v}$, construct neighbor messages through $<$product, relationship, neighbor product$>$, and perform message aggregation. As shown in the formulas (\ref{eqn:105})-(\ref{eqn:106}).
\begin{equation}
	m_{v \leftarrow w}=f_{c}\left(h_{v}, e_{w v}, h_{w}\right)\label{eqn:105}
\end{equation}
\begin{equation}
	m_{v}=\frac{1}{\left|\mathcal{N}_{v}\right|} \sum_{w \in \mathcal{N}_{v}} m_{v \leftarrow w}\label{eqn:106}
\end{equation}
Among them, $f_{c}\left(\cdot\right)$ is the message constructor, $h_{v}$ is the query product embedding, and $h_{w}$ is the neighbor product embedding.
\subsubsection{Relationship strength}
\
\newline The strength of the complementary relationship between different pairs of products is different. For example, the complementarity between bread and milk is stronger than the complementarity between bread and lunch meat because 10.5 times more people buy bread and milk than bread and lunch meat. In order to bring a better shopping experience, it is reasonable to give priority to recommending products that are closely related to the query product. Therefore, the strength of the relationship is important for complementary recommendations. Most literature regards product relationship prediction as a classification task, and then uses Hinge loss\citep{hao2020p,yan2022personalized} or cross-entropy loss function\citep{liu2020decoupled,zhang2019identifying,liu2021item,yang2022inferring} to classify the positive and negative examples. There is no pay attention to the strength of the relationship between products. \citep{DBLP:conf/aaai/ChenHXFLSYQ23} solves this problem and establishes an auxiliary loss function based on the triplet loss between the query product and two complementary products, as shown in the formula (\ref{eqn:107}), to distinguish which product is more relevant to the query product.
\begin{equation}
	\mathcal{L}^{\prime}\left(z_{i}^{c}\right)=\mathrm{I} * \max \left\{0, s\left(z_{i}^{c}, z_{\text {pos }_{1}}^{c}\right)-s\left(z_{i}^{c}, z_{\text {pos }_{2}}^{c}\right)+\Delta^{\prime}\right\}\label{eqn:107}
\end{equation}
\begin{equation}
	\mathrm{I}=\left\{\begin{array}{ll}
		1 & \text { if } r^{c}\left(v_{i}, v_{\text {pos }_{1}}\right)<r^{c}\left(v_{i}, v_{\text {pos }_{2}}\right) \\
		-1 & \text { else },
	\end{array}\right.\label{eqn:108}
\end{equation}
\begin{equation}
	\Delta^{\prime}=\gamma *\left(r^{c}\left(v_{i}, v_{p o s_{2}}\right)-r^{c}\left(v_{i}, v_{\text {pos }_{1}}\right)\right)\label{eqn:109}
\end{equation}
Among them, $v_{p o s_{1}}, v_{p o s_{2}}$ are two complementary products of product $v_{i}$. $r^{c}$ is a piecewise function of the strength of the complementary relationship, which divides the complementary relationship into different levels through predetermined strength values.
\section{Comparative analysis of complementary recommendation models}
As shown in Table \ref{tab:my-table}, the research issues of the complementary recommendation model are classified and introduced in detail.
\begin{table*}[]
	\centering
	\caption{Classification comparison of complementary recommendation models}
	\label{tab:my-table}
	\resizebox{\textwidth}{!}{%
		\begin{tabular}{|c|c|c|c|}
			\hline
			research problem                 & literature                                                      & challenge                            & Solution                                                     \\ \hline
			\multirow{2}{*}{diversity}       & \citep{hao2020p}                                & weigh accuracy and diversity         & category diversity                                           \\ \cline{2-4} 
			& \citep{ma2023personalized}                      & personalized diversity               & user classification                                          \\ \hline
			\multirow{7}{*}{personalization} & \citep{wan2018representing}                     & shopping basket personalization      & triple2vec gets user embedding representation                \\ \cline{2-4} 
			& \citep{ma2023personalized}                      & shopping basket personalization      & triple2vec gets user embedding representation                \\ \cline{2-4} 
			& \citep{trofimov2018inferring}                   & shopping basket personalization      & average product embedding                                    \\ \cline{2-4} 
			& \citep{entezari2021tensor} & shopping basket personalization      & tensor decomposition to obtain user embedding representation \\ \cline{2-4} 
			& \citep{yan2022personalized}                     & personalized recommendations         & transformer encoder                                          \\ \cline{2-4} 
			& \citep{zhang2021learning}                       & personalized sequence recommendation & transformer encoder                                          \\ \cline{2-4} 
			& \citep{zhang2018quality}                        & product quality personalization      & bayesian inference                                           \\ \hline
			\multirow{4}{*}{cold start}      & \citep{DBLP:conf/www/BibasSJ23}                           & new product recommendation           & CycleGAN                                                     \\ \cline{2-4} 
			& \citep{wang2018path}                            & new product recommendations          & use product categories as product embeddings                 \\ \cline{2-4} 
			& \citep{DBLP:conf/recsys/Papso23}                & long tail product recommendations    & transfer learning                                            \\ \cline{2-4} 
			& \citep{xu2020knowledge}                         & new product recommendations          & use product contextual information                           \\ \hline
			scenario based                   & \citep{kang2019complete}                        & real world scene effects             & calculate product and scene compatibility                    \\ \hline
			\multirow{2}{*}{interpretability}                 & \citep{yang2022inferring}                       & recommended explainability           & knowledge graph, reinforcement learning                      \\ \cline{2-4}
			& \citep{DBLP:journals/corr/abs-2305-11480}       & recommended explainability           & large language model                                         \\ \hline
			\multirow{2}{*}{data noise}                       & \citep{xu2020knowledge}                         & complementary labels are noisy       & multi-task learning, contextual knowledge                    \\  \cline{2-4}
			& \citep{ma2021neat}                              & complementary labels are noisy       & gaussian embedding                                           \\ \hline
		\end{tabular}%
	}
\end{table*}
\subsection{Diversity}
\begin{figure}[htb]
	\centering
	\includegraphics[scale=1]{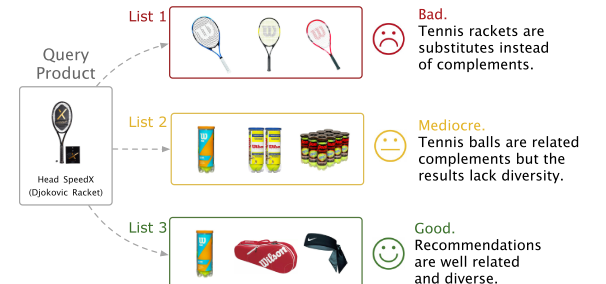}
	\caption{Complementary recommendation diversity}\label{fig:fig7}
\end{figure}

As shown in Figure \ref{fig:fig7}, given a "query product" tennis racket, three sets of complementary recommendation lists are given. List 1 shows three other similar tennis rackets; List 2 shows three tennis balls; List 3 shows a tennis ball, a racket cover and a hair tie. In general, we think List 1 is more of an alternative recommendation. While both List 2 and List 3 can be considered reasonable complementary recommendations, we believe List 3 is the better choice because List 3 lists three different types of products that together better meet the customer's needs for tennis. This example illustrates that an ideal complementary product recommendation should consider both relevance and diversity to meet customer needs. \citep{hao2020p,ma2023personalized}realize the problem of product diversity for complementary recommendations.
\newline \indent \citep{hao2020p}proposed the P-Companion framework, which uses encoder-decoder to calculate the category of complementary products according to the category of the query product, such as the formula (\ref{eqn:51})-(\ref{eqn:52}), and then select the most complementary product in each complementary category for recommendation. Achieving diversity of strictly complementary recommended products.
\begin{equation}
	h=\operatorname{Dropout}\left(\operatorname{ReLU}\left(\phi_{w_{i}} W^{(4)}+b^{(4)}\right)\right)\label{eqn:51}
\end{equation}
\begin{equation}
	\gamma_{w_{i}}=h W^{(5)}+b^{(5)}\label{eqn:52}
\end{equation}
Among them, $W^{(4)}$ and $W^{(5)}$ are the encoder and decoder weights respectively.
\newline \indent \citep{hao2020p}recommends a variety of complementary products to all users. However, not all users like this recommendation method. \citep{ma2023personalized}divides users into two types, exploratory users and traditional users, and uses different recommendation methods for different types of users.
\newline \indent When the user is an exploratory user, he may prefer to see diverse products that are complementary to the query product, because the user wants to explore and rerank the recommended candidate complementary products to include more diverse products. As shown in the following formula:
\begin{equation}
	r_{j}=\arg \max _{r_{i} \in R \backslash R_{d}} \underbrace{\alpha S_{q, r_{i}}}_{\text {complementarity }}+\underbrace{(1-\alpha)\left(\log \operatorname{det}\left(\mathbf{L}_{R_{d}+\left[r_{i}\right]}\right)-\log \operatorname{det}\left(\mathbf{L}_{R_{d}}\right)\right)}_{\text {increment of diversification }}\label{eqn:53}
\end{equation}
Where $R_{t-1,d}+[r_{j}]$ means that the newly selected recommended product $r_{j}$ by diversified reranking is inserted at the end of the current list of products $R_{t-1,d}$. The weight $\alpha$ controls the amount of diversity introduced into the reranked product list. Each selected item $r_{j}$ can maximize the combination score of diversity and complementarity. Compared with the original recommended product list $R$, the reranked product list $R_{d}$ will present more diverse recommendations toward the top, where products are simply ranked by score $S_{q, r_{i}}$ in descending order.
\newline \indent When the user is a traditional user, he may prefer the system to recommend less diverse products that are complementary to the query product, because he likes classic complementary combinations. As shown in the following formula:
\begin{equation}
	r_{h}=\arg \max _{r_{i} \in\left(R+R_{x}\right) \backslash R_{s}} \underbrace{\beta S_{q, r_{i}}}_{\text {complementarity }} +  \underbrace{(1-\beta)\left(\log \operatorname{det}\left(\mathbf{L}_{R_{s}+\left[r_{i}\right]}^{\prime}\right)-\log \operatorname{det}\left(\mathbf{L}_{R_{s}}^{\prime}\right)\right)}_{\text {increment of similarity between recommendations }}\label{eqn:54}
\end{equation}
Among them, the parameter $\beta$ is used to control the degree of similarity between recommended products.
\subsection{Personalization}
\begin{figure}[htb]
	\centering
	\includegraphics[scale=0.8]{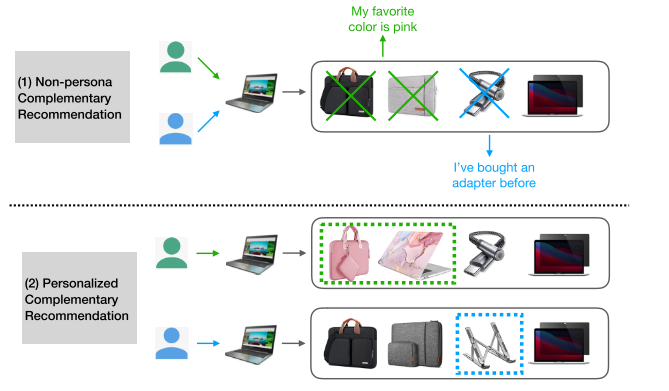}
	\caption{Non-personalized recommendations and personalized recommendations}\label{fig:fig3}
\end{figure}

If complementary recommendation only considers the general complementary relationship between products, it may meet the needs of most users, but it cannot meet the needs of users who have their own preferences. As shown in Figure\ref{fig:fig3}, when the given query product is a laptop, non-personalized recommendations recommend the same complementary products to all users. If users have different preferences for colors or have purchased the recommended product before, then the recommendation effect is not ideal. Therefore, user preferences should be considered to recommend products suitable for different users.
\newline \indent \citep{rakesh2019linked}combines Probabilistic Matrix Factorization (PMF) with Linked Variational Autoencoders (LVA) to create CLVA that captures the implicit relationship between users and products.\citep{wan2018representing,entezari2021tensor,ma2023personalized,trofimov2018inferring}is to recommend shopping basket products. \citep{wan2018representing,ma2023personalized}uses the triple2vec method to obtain product complementarity and user preferences.
\citep{wan2018representing}considers two recommendation scenarios: personalized next shopping basket product recommendation and recommended products in the shopping basket. personalized next shopping basket item recommendation: Given a user, when recommending products for the next shopping basket, the product embedding of the given product is replaced by the average embedding of all products in the user's previous shopping basket, and then the new user can be obtained the preference score $s_{i,u}$, the purchase probability is:
\begin{equation}
	p_{i, u}=\frac{\exp \left(s_{i, u}\right)}{\sum_{i^{\prime}} \exp \left(s_{i^{\prime}, u}\right)}\label{eqn:18}
\end{equation}
\newline \indent recommended products in the shopping basket: Given the products in the current shopping basket, when recommending products to be added to the same basket, replace product embeddings with the average embedding of all products in the given product set. \citep{wan2018representing}believes that many users have their own must-buy products. This situation can be measured by the frequency of users purchasing products, but will not be captured by low-dimensional product representation and user representation. Therefore, the algorithm adaLoyal is proposed to adaptively combine these two components and estimate users' product loyalty over time. The following is the probability of the end user purchasing the product:
\begin{tiny}
	\begin{equation}
		P\left(C_{i, u}^{(t)}=1\right)=P\left(L_{i, u}\right) P\left(C_{i, u}^{(t)}=1 \mid L_{i, u}\right)+P\left(\neg L_{i, u}\right) P\left(C_{i, u}^{(t)}=1 \mid \neg L_{i, u}\right)\label{eqn:19}
	\end{equation}
\end{tiny}
Among them, $L_{i, u}$ represents the user's loyalty index to the product, and $P\left(C_{i, u}^{(t)}=1 \mid L_{i, u}\right)$ is user purchase frequency model, $P\left(C_{i, u}^{(t)}=1 \mid \neg L_{i, u}\right)$ is the user product representation model.
\newline \indent \citep{entezari2021tensor}uses tensor decomposition technology to obtain the embedding representation $\underline{R}_{k}$ of user $k$, $\underline{R}_{k}$ captures user $k $ interactions between purchased products, so use $\underline{R}_{k}$ to adjust the complementarity scores between products.
\newline \indent \citep{zhang2021learning,yan2022personalized}based on the user's purchase sequence, the product purchase sequence is sent to the transformer encoder, and the user feature vector is learned to express user preferences, as shown in the formula (\ref{eqn:21}), use the first embedding of the transformer hidden state as the contextual user embedding $u$.
\begin{equation}
	u=transformer(t_{1},t_{2},...,t_{n})\label{eqn:21}
\end{equation}
\newline \indent \citep{zhang2018quality}pays more attention to the quality of complementary products. For example, there are dozens of Apple phone cases that are complementary products to Apple phones. Which one will the user choose? This means getting the user's quality expectations and then making personalized recommendations. Bayesian inference based on user ratings is mainly used to model the underlying quality of complementary products. As shown in the formula (\ref{eqn:22}), it is the user's quality expectation for the complementary products of product $I_{i}$.
\begin{equation}
	\mathbb{E}\left(\theta_{i} \mid q_{i 1}, q_{i 2} \ldots q_{i\left|I_{i}\right|}\right)=\frac{\sum_{k=1}^{\left|I_{i}\right|} q_{i k}+1}{\left|I_{i}\right|+2}\label{eqn:22}
\end{equation}
\subsection{Scenario based}
\begin{figure}[htb]
	\centering
	\includegraphics[scale=0.8]{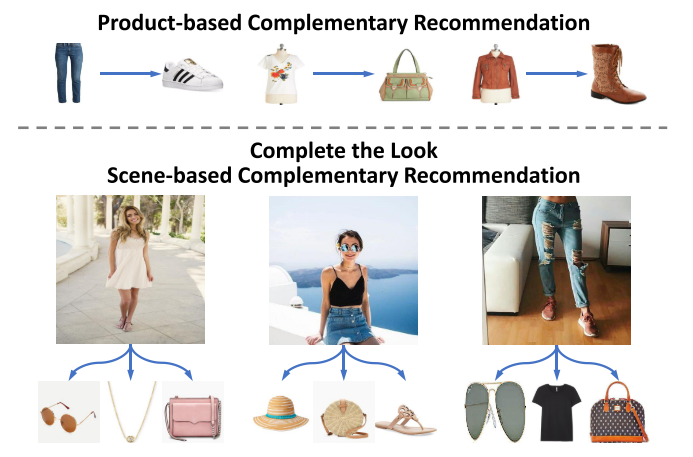}
	\caption{Comparative analysis of product-based complementary recommendations and scenario-based complementary recommendations}\label{fig:fig22}
\end{figure}
As shown in Figure \ref{fig:fig22}, in the field of clothing recommendation, product-based complementary recommendations only predict the compatibility between two products based on product images, ignoring real-world scene images. Scene images provide key information such as the user's body type or season. Therefore, complementary clothing recommendations can be made based on scene images.
\newline \indent \citep{kang2019complete}uses the current scene image and candidate products to calculate the compatibility of the product and the scene, thereby making more accurate recommendations.
\subsection{Cold start}
The cold start problem refers to what to do to recommend a product for a newly registered user or how to recommend a newly stocked product to a user who likes it. \citep{DBLP:conf/www/BibasSJ23,wang2018path,DBLP:conf/recsys/Papso23,xu2020knowledge}mainly proposes a solution to this problem. \citep{DBLP:conf/www/BibasSJ23}for products without interactive data, the Cycle Generative Adversarial Network (CycleGAN) is used. The model is divided into two parts: a classifier and a reconstructor. The classifier receives product embedding $v_{i}$ and category $c$ as input, and outputs the probability that product $i$ belongs to category $c$; the reconstructor inputs the original category of the product and the representation of projecting the product into a specific category, reconstruct the original embedding of the product. Finally, Euclidean distance is used to measure the complementary distance of two product. \citep{wang2018path}believes that when products are related, the categories to which the products belong are also relevant. Therefore, for new products, the feature representation of the category to which they belong is used. The class feature vector looks like this:
\begin{equation}
	v_{c, r}=\frac{\sum_{i \in c} v_{i, r}}{N_{c}}, v_{c, r}^{\prime}=\frac{\sum_{i \in c} v_{i, r}^{\prime}}{N_{c}}\label{eqn:61}
\end{equation}
Among them, $v_{c}$ and $v_{c}^{\prime}$ are the query vector and complementary vector of category $c$ respectively; $r$ is the relationship type; $N_{c}$ is the number of products in category $c$.
\newline \indent \citep{DBLP:conf/recsys/Papso23}using transfer learning to solve the cold start problem. The main idea of transfer learning is to train a general model on a large amount of data and then apply (i.e. transfer) the learned knowledge to a more specific task for which little data is available. First, a product universal embedding space is established based on product characteristics. When a new product or even a new category appears, transfer learning is used to project the new product into the product universal embedding space. \citep{xu2020knowledge}fuses the contextual information of the product into the product embedding representation. As shown in the formula (\ref{eqn:61}), the product embedding can be inferred using only the contextual embedding of the trained cold-start product.
\begin{equation}
	\hat{\mathbf{z}}=\arg \max _{\mathbf{z} \in \mathbb{R}^{P}} \prod_{w \in \mathbf{w}_{i_{0}}} p(w \mid \mathbf{z})\label{eqn:62}
\end{equation}
\subsection{Interpretability}
\begin{figure}[htb]
	\centering
	\includegraphics[scale=0.6]{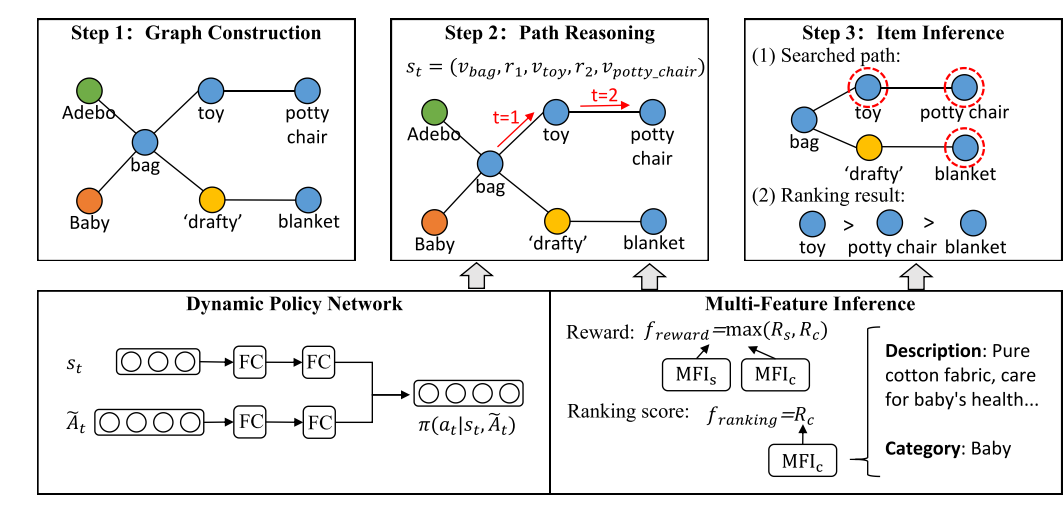}
	\caption{Knowledge-aware path reasoning flow graph}\label{fig:fig80}
\end{figure}
\citep{yang2022inferring}proposed a knowledge-aware path reasoning method, which constructs the co-purchase relationship of products and product properties into a knowledge graph, and uses the idea of reinforcement learning to perform reasoning, making the reasoning of complementary products interpretable.As shown in Figure \ref{fig:fig80}.
\newline \indent \citep{DBLP:journals/corr/abs-2305-11480}uses a large language model to generate a complementary product list for the product. For example, entering "digital camera" can generate 1) camera lens 2) battery 3) memory card, And when asking the big language model "Why can memory cards be purchased together with digital cameras?", a detailed explanation can be given: "Memory cards provide storage space for photos taken by digital cameras." This allows users to better understand why this product is recommended. The formula looks like this:
\begin{equation}
	P_{\theta}(s \mid x)=\prod_{i=1}^{|s|} P_{\theta}\left(s_{i} \mid s_{0: i-1}, \operatorname{prompt}(x)\right)\label{eqn:61}
\end{equation}
Where $s$ is the generated complementary product list.
\subsection{Data noise}
\begin{figure}[htb]
	\centering
	\includegraphics[scale=0.8]{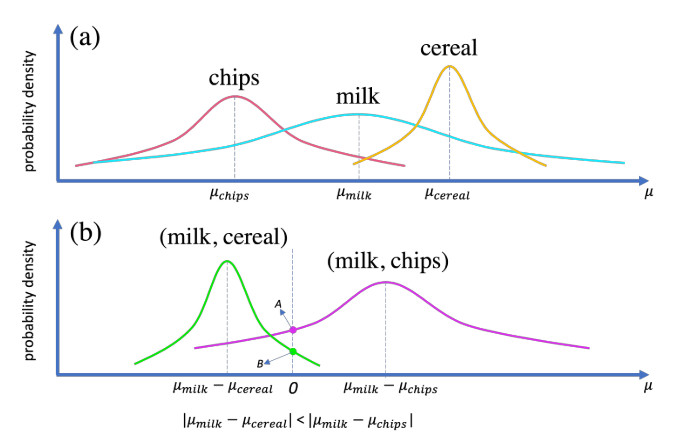}
	\caption{Examples of Gaussian distributions for milk, cereal, and chips}\label{fig:fig6}
\end{figure}
Most existing complementary recommendation systems use co-purchase information of products as complementary relationships. However, co-purchased products are not necessarily complementary. For example, a customer may often purchase bananas and bottled water in the same transaction, but the two products are not complementary. Therefore, directly using co-purchase signals as labels will affect model performance. On the other hand, if the labels of the evaluation do not reflect true complementary relationships, then the model evaluation is not trustworthy. \citep{xu2020knowledge,ma2021neat}proposed solutions. \citep{xu2020knowledge}encodes contextual knowledge into product representations through multi-task learning to alleviate the sparsity problem. \citep{ma2021neat}learns product representation using Gaussian embeddings. Co-purchase of two products is modeled as a Gaussian distribution, where the mean represents the co-purchase from complementary correlation and the covariance represents the co-purchase from noise. Figure \ref{fig:fig6} shows an example of daily shopping. In Figure \ref{fig:fig6}(a), milk has the largest variance among the three commodities because it is usually purchased by many customers and is likely to be purchased together with other products that have no complementary relationship. Since the co-purchase behavior of cereal and milk is stable, cereal has the smallest variance. The variance of chipss is medium because it has some stable combinations like chips with ketchup, and users may also buy chips alone as a snack before checking out, which makes its variance relatively large. In Figure \ref{fig:fig6}(b), the Gaussian distribution of their complementary relationship is shown, and when milk is used as the query product, at point A (milk, chips) and point b (milk, cereal) their probability densities at 0 are highlighted respectively. Due to the difference in variance, the Gaussian distribution for buying (milk,cereal) has a smaller variance than the Gaussian distribution for buying (milk,chips). Even though there may be more observed co-purchase records between milk and chips than between milk and cereal, the model can still be improved by comparing $|\mu_{milk}-\mu_{cereal}|$ and $|\mu_ { milk}-\mu_{chips}|$ to capture the correct ordering of complementary relationships.
\section{Comparative analysis of experimental results}
\begin{table*}[htb]
	\centering
	\caption{Comparative analysis of datasets and complementary label in complementary recommendation}
	\label{table3}
	\resizebox{0.8\textwidth}{!}{%
		\begin{tabular}{|c|c|c|c|}
			\hline
			literature                                           & code                                                                                                                  & complementary label           & datasets     \\ \hline
			\citep{DBLP:conf/aaai/ChenHXFLSYQ23} &           \textbackslash{}                                                                                                            & buy together                  & Amazon        \\ \hline
			\citep{DBLP:conf/www/BibasSJ23}                & { https://fb.me/cgan\_ccomplementary\_item\_recommendation}                                 & buy together                  & Amazon        \\ \hline
			\citep{DBLP:conf/recsys/Papso23}     & https://github.com/r-papso/carca-replication                                                                          & LLM Automatic data annotation & Amazon        \\ \hline
			\citep{yan2022personalized}          &                     \textbackslash{}                                                                                                  & buy together                  & other         \\ \hline
			\citep{wu2022towards}                &                                     \textbackslash{}                                                                                  & buy together-view together    & Amazon        \\ \hline
			\citep{DBLP:conf/cikm/ZhouWHZMD22}   & https://github.com/wt-tju/DHGAN                                                                                       & buy together                  & Amazon        \\ \hline
			\citep{yang2022inferring}            & https://gitee.com/yangzijing\_flower/kapr/tree/master                                                                 & buy together                  & Amazon        \\ \hline
			\citep{zhang2021learning}            &                                                     \textbackslash{}                                                                  & buy together                  & Amazon        \\ \hline
			\citep{liu2021item}                  &  https://github.com/wwliu555/IRGNN\_TNNLS\_2021 & buy together                  & Amazon+Taobao \\ \hline
			\citep{liu2020decoupled}             & https://github.com/guyulongcs/CIKM2020\_DecGCN                                                                        & buy together                  & Amazon+JD     \\ \hline
			\citep{hao2020p}                     &                                                                     \textbackslash{}                                                  & buy together-view together    & Amazon        \\ \hline
			\citep{rakesh2019linked}             & https://github.com/VRM1/WSDM19                                                                                        & buy together                  & Amazon        \\ \hline
			\citep{zhang2019identifying}         &                                       \textbackslash{}                                                                                & buy together                  & Amazon        \\ \hline
			\citep{wang2018path}                 &                                                       \textbackslash{}                                                                & buy together                  & Amazon        \\ \hline
			\citep{zhao2017deep}                 &                                                                       \textbackslash{}                                                & buy together+manual marking   & Amazon+Taobao \\ \hline
			\citep{mcauley2015inferring}         & https://github.com/abeermohamed1/Recommender-System                                                                   & buy together                  & Amazon        \\ \hline
			\citep{DBLP:journals/corr/abs-2401-02130}         & https://github.com/luohaitong/SComGNN                                                                   & buy together                  & Amazon        \\ \hline
		\end{tabular}%
	}
\end{table*}
\subsection{Datasets}
\subsubsection{Dataset name}
\
\newline As shown in Table \ref{table3}, the complementary recommended datasets are mainly as follows:
\newline \indent \textbf{Amazon.} The dataset includes reviews and metadata information from 142.8 million products on Amazon from May 1996 to July 2014.
\newline \indent \textbf{Instacart.} The announcement data set released by the Kaggle competition in 2017 contains 3 million purchase orders from more than 200,000 users.
\newline \indent \textbf{JD.} Large industrial dataset collected from JD.com.
\newline \indent \textbf{TaoBao.} From Taobao, provided by Alibaba Group. \citep{zhao2017deep} Using clothing category data, there are approximately 406k clothing compatibility relationships, covering 61k products. The compatibility relationships are manually marked by clothing matching experts. \citep{liu2021item} uses user behavior data.
\subsubsection{Complementary label}
\
\newline As shown in Table \ref{table3}, \citep{DBLP:conf/aaai/ChenHXFLSYQ23,DBLP:conf/www/BibasSJ23,yan2022personalized,zhang2021learning,zhao2017deep,mcauley2015inferring,DBLP:conf/cikm/ZhouWHZMD22,liu2020decoupled,liu2021item,rakesh2019linked,wang2018path,zhang2019identifying,zhang2018quality}use co-purchase products as complementary products. \citep{DBLP:conf/recsys/Papso23}use hint-based LLM like ChatGPT to automate complementary data annotation. \citep{wu2022towards,hao2020p} believes that buy together data overlaps with view together data, and uses (buy together-view together) as complementary product label.
\subsection{Model training}
\subsubsection{loss function}
\
\newline The loss functions used by the complementary recommendation model are mainly as follows:
\newline \indent \textbf{Cross entropy loss function.} The standard form of the cross-entropy loss function is as follows:
\begin{equation}
	C=-\frac{1}{n} \sum_{x}[y \ln a+(1-y) \ln (1-a)]\label{eqn:91}
\end{equation}
Where $x$ represents the sample, $y$ represents the actual label, $a$ represents the predicted output, and $n$ represents the total number of samples. Can be used in binary classification and multi-classification tasks.
\newline \indent \textbf{Hinge loss function.} The standard form of the hinge loss function is as follows:
\begin{equation}
	L(y, f(x))=\max (0,1-y f(x))\label{eqn:92}
\end{equation}
If it is classified correctly, the loss is 0, otherwise the loss is $1-y f(x)$.
\subsubsection{Optimization}
\
\newline The optimization methods used by the complementary recommendation model are mainly as follows: Stochastic gradient descent method (SGD) \citep{zhao2017deep,ma2021neat}. Adaptive motion estimation algorithm (Adam) \citep{zhang2021learning,DBLP:conf/recsys/Papso23,liu2020decoupled,liu2021item,zhang2019identifying}.Expectation maximization (EM)\citep{zhao2017recommending,mcauley2015inferring}.
\subsection{Evaluation}
\begin{table*}[htb]
	\centering
	\caption{Comparative analysis of experimental results}
	\label{tab:my-table1}
	\resizebox{0.9\textwidth}{!}{%
		\begin{tabular}{|c|c|ccccccccccccccccccc|}
			\hline
			& \multicolumn{1}{c|}{}                                & \multicolumn{19}{c|}{evaluation}                                          \\ \cline{3-21} 
			\multirow{-2}{*}{databases}           & \multicolumn{1}{c|}{\multirow{-2}{*}{literature}}    & \multicolumn{1}{c|}{Accuracy} & \multicolumn{1}{c|}{Precision} & \multicolumn{1}{c|}{AUC}    & \multicolumn{1}{c|}{Recall@30} & \multicolumn{1}{c|}{Recall} & \multicolumn{1}{c|}{NDGC@3} & \multicolumn{1}{c|}{NDGC@5} & \multicolumn{1}{c|}{NDGC@10} & \multicolumn{1}{c|}{NDGC@30} & \multicolumn{1}{c|}{NDGC} & \multicolumn{1}{c|}{HR@1} & \multicolumn{1}{c|}{HR@3} & \multicolumn{1}{c|}{HR@5} & \multicolumn{1}{c|}{HR@10} & \multicolumn{1}{c|}{HR@30} & \multicolumn{1}{c|}{HR@50} & \multicolumn{1}{c|}{MRR}   & \multicolumn{1}{c|}{MRR@10} & F1     \\ \hline
			&\citep{DBLP:conf/aaai/ChenHXFLSYQ23} & \multicolumn{1}{c|}{}         & \multicolumn{1}{c|}{}          & \multicolumn{1}{c|}{}       & \multicolumn{1}{c|}{0.9562}                                 & \multicolumn{1}{c|}{}                                    & \multicolumn{1}{c|}{}                                    & \multicolumn{1}{c|}{}                                    & \multicolumn{1}{c|}{}                                     & \multicolumn{1}{c|}{0.8371}                               & \multicolumn{1}{c|}{}                                  & \multicolumn{1}{c|}{}                                  & \multicolumn{1}{c|}{}                                  & \multicolumn{1}{c|}{}                                  & \multicolumn{1}{c|}{}      & \multicolumn{1}{c|}{}      & \multicolumn{1}{c|}{}      & \multicolumn{1}{c|}{}      & \multicolumn{1}{c|}{}                                    &        \\ \cline{2-21} 
			& \citep{yang2022inferring}            & \multicolumn{1}{c|}{}         & \multicolumn{1}{c|}{}          & \multicolumn{1}{c|}{}       & \multicolumn{1}{c|}{}                                       & \multicolumn{1}{c|}{}                                    & \multicolumn{1}{c|}{}                                    & \multicolumn{1}{c|}{}                                    & \multicolumn{1}{c|}{}                                     & \multicolumn{1}{c|}{}                                     & \multicolumn{1}{c|}{}                                  & \multicolumn{1}{c|}{}                                  & \multicolumn{1}{c|}{}                                  & \multicolumn{1}{c|}{}                                  & \multicolumn{1}{c|}{0.89}                               & \multicolumn{1}{c|}{0.95}                               & \multicolumn{1}{c|}{0.95}                               & \multicolumn{1}{c|}{}      & \multicolumn{1}{c|}{}                                    &        \\ \cline{2-21} 
			& \citep{liu2020decoupled}             & \multicolumn{1}{c|}{}         & \multicolumn{1}{c|}{}          & \multicolumn{1}{c|}{}       & \multicolumn{1}{c|}{}                                       & \multicolumn{1}{c|}{}                                    & \multicolumn{1}{c|}{}                                    & \multicolumn{1}{c|}{}                                    & \multicolumn{1}{c|}{0.593}                                & \multicolumn{1}{c|}{}                                     & \multicolumn{1}{c|}{}                                  & \multicolumn{1}{c|}{}                                  & \multicolumn{1}{c|}{}                                  & \multicolumn{1}{c|}{}                                  & \multicolumn{1}{c|}{0.716}                              & \multicolumn{1}{c|}{}                                   & \multicolumn{1}{c|}{}                                   & \multicolumn{1}{c|}{}      & \multicolumn{1}{c|}{0.471}                               &        \\ \cline{2-21} 
			\multirow{-4}{*}{Amazon-Beauty}       & \citep{zhang2019identifying}         & \multicolumn{1}{c|}{0.859}    & \multicolumn{1}{c|}{}          & \multicolumn{1}{c|}{}       & \multicolumn{1}{c|}{}                                       & \multicolumn{1}{c|}{}                                    & \multicolumn{1}{c|}{}                                    & \multicolumn{1}{c|}{}                                    & \multicolumn{1}{c|}{}                                     & \multicolumn{1}{c|}{}                                     & \multicolumn{1}{c|}{}                                  & \multicolumn{1}{c|}{}                                  & \multicolumn{1}{c|}{}                                  & \multicolumn{1}{c|}{}                                  & \multicolumn{1}{c|}{}                                   & \multicolumn{1}{c|}{}                                   & \multicolumn{1}{c|}{}                                   & \multicolumn{1}{c|}{}      & \multicolumn{1}{c|}{}                                    & 0.8116 \\ \hline
			& \citep{DBLP:conf/aaai/ChenHXFLSYQ23} & \multicolumn{1}{c|}{}         & \multicolumn{1}{c|}{}          & \multicolumn{1}{c|}{}       & \multicolumn{1}{c|}{0.9144}                                 & \multicolumn{1}{c|}{}                                    & \multicolumn{1}{c|}{}                                    & \multicolumn{1}{c|}{}                                    & \multicolumn{1}{c|}{}                                     & \multicolumn{1}{c|}{0.7425}                               & \multicolumn{1}{c|}{}                                  & \multicolumn{1}{c|}{}                                  & \multicolumn{1}{c|}{}                                  & \multicolumn{1}{c|}{}                                  & \multicolumn{1}{c|}{}      & \multicolumn{1}{c|}{}      & \multicolumn{1}{c|}{}      & \multicolumn{1}{c|}{}      & \multicolumn{1}{c|}{}                                    &        \\ \cline{2-21} 
			& \citep{DBLP:conf/cikm/ZhouWHZMD22}   & \multicolumn{1}{c|}{}         & \multicolumn{1}{c|}{}          & \multicolumn{1}{c|}{}       & \multicolumn{1}{c|}{}                                       & \multicolumn{1}{c|}{}                                    & \multicolumn{1}{c|}{}                                    & \multicolumn{1}{c|}{}                                    & \multicolumn{1}{c|}{0.635}                                & \multicolumn{1}{c|}{}                                     & \multicolumn{1}{c|}{}                                  & \multicolumn{1}{c|}{}                                  & \multicolumn{1}{c|}{}                                  & \multicolumn{1}{c|}{}                                  & \multicolumn{1}{c|}{0.78}                               & \multicolumn{1}{c|}{}                                   & \multicolumn{1}{c|}{}                                   & \multicolumn{1}{c|}{}      & \multicolumn{1}{c|}{0.59}                                &        \\ \cline{2-21} 
			& \citep{yang2022inferring}            & \multicolumn{1}{c|}{}         & \multicolumn{1}{c|}{}          & \multicolumn{1}{c|}{}       & \multicolumn{1}{c|}{}                                       & \multicolumn{1}{c|}{}                                    & \multicolumn{1}{c|}{}                                    & \multicolumn{1}{c|}{}                                    & \multicolumn{1}{c|}{}                                     & \multicolumn{1}{c|}{}                                     & \multicolumn{1}{c|}{}                                  & \multicolumn{1}{c|}{}                                  & \multicolumn{1}{c|}{}                                  & \multicolumn{1}{c|}{}                                  & \multicolumn{1}{c|}{0.66}                               & \multicolumn{1}{c|}{0.9}                                & \multicolumn{1}{c|}{0.93}                               & \multicolumn{1}{c|}{}      & \multicolumn{1}{c|}{}                                    &        \\ \cline{2-21} 
			& \citep{liu2021item}                  & \multicolumn{1}{c|}{0.7413}   & \multicolumn{1}{c|}{0.757}     & \multicolumn{1}{c|}{0.9032} & \multicolumn{1}{c|}{}                                       & \multicolumn{1}{c|}{0.7304}                              & \multicolumn{1}{c|}{}                                    & \multicolumn{1}{c|}{}                                    & \multicolumn{1}{c|}{}                                     & \multicolumn{1}{c|}{}                                     & \multicolumn{1}{c|}{}                                  & \multicolumn{1}{c|}{}                                  & \multicolumn{1}{c|}{}                                  & \multicolumn{1}{c|}{}                                  & \multicolumn{1}{c|}{}                                   & \multicolumn{1}{c|}{}                                   & \multicolumn{1}{c|}{}                                   & \multicolumn{1}{c|}{}      & \multicolumn{1}{c|}{}                                    &        \\ \cline{2-21} 
			& \citep{liu2020decoupled}             & \multicolumn{1}{c|}{}         & \multicolumn{1}{c|}{}          & \multicolumn{1}{c|}{}       & \multicolumn{1}{c|}{}                                       & \multicolumn{1}{c|}{}                                    & \multicolumn{1}{c|}{}                                    & \multicolumn{1}{c|}{}                                    & \multicolumn{1}{c|}{0.429}                                & \multicolumn{1}{c|}{}                                     & \multicolumn{1}{c|}{}                                  & \multicolumn{1}{c|}{}                                  & \multicolumn{1}{c|}{}                                  & \multicolumn{1}{c|}{}                                  & \multicolumn{1}{c|}{0.583}                              & \multicolumn{1}{c|}{}                                   & \multicolumn{1}{c|}{}                                   & \multicolumn{1}{c|}{}      & \multicolumn{1}{c|}{0.311}                               &        \\ \cline{2-21} 
			& \citep{hao2020p}                     & \multicolumn{1}{c|}{}         & \multicolumn{1}{c|}{}          & \multicolumn{1}{c|}{}       & \multicolumn{1}{c|}{}                                       & \multicolumn{1}{c|}{}                                    & \multicolumn{1}{c|}{}                                    & \multicolumn{1}{c|}{}                                    & \multicolumn{1}{c|}{}                                     & \multicolumn{1}{c|}{}                                     & \multicolumn{1}{c|}{}                                  & \multicolumn{1}{c|}{0.145}                             & \multicolumn{1}{c|}{0.17}                              & \multicolumn{1}{c|}{}                                  & \multicolumn{1}{c|}{0.187}                              & \multicolumn{1}{c|}{}                                   & \multicolumn{1}{c|}{}                                   & \multicolumn{1}{c|}{}      & \multicolumn{1}{c|}{}                                    &        \\ \cline{2-21} 
			& \citep{rakesh2019linked}             & \multicolumn{1}{c|}{0.9118}   & \multicolumn{1}{c|}{}          & \multicolumn{1}{c|}{}       & \multicolumn{1}{c|}{}                                       & \multicolumn{1}{c|}{}                                    & \multicolumn{1}{c|}{}                                    & \multicolumn{1}{c|}{}                                    & \multicolumn{1}{c|}{}                                     & \multicolumn{1}{c|}{}                                     & \multicolumn{1}{c|}{}                                  & \multicolumn{1}{c|}{}                                  & \multicolumn{1}{c|}{}                                  & \multicolumn{1}{c|}{}                                  & \multicolumn{1}{c|}{}                                   & \multicolumn{1}{c|}{}                                   & \multicolumn{1}{c|}{}                                   & \multicolumn{1}{c|}{}      & \multicolumn{1}{c|}{}                                    &        \\ \cline{2-21} 
			& \citep{zhang2019identifying}         & \multicolumn{1}{c|}{0.9219}   & \multicolumn{1}{c|}{}          & \multicolumn{1}{c|}{}       & \multicolumn{1}{c|}{}                                       & \multicolumn{1}{c|}{}                                    & \multicolumn{1}{c|}{}                                    & \multicolumn{1}{c|}{}                                    & \multicolumn{1}{c|}{}                                     & \multicolumn{1}{c|}{}                                     & \multicolumn{1}{c|}{}                                  & \multicolumn{1}{c|}{}                                  & \multicolumn{1}{c|}{}                                  & \multicolumn{1}{c|}{}                                  & \multicolumn{1}{c|}{}                                   & \multicolumn{1}{c|}{}                                   & \multicolumn{1}{c|}{}                                   & \multicolumn{1}{c|}{}      & \multicolumn{1}{c|}{}                                    & 0.9032 \\ \cline{2-21} 
			& \citep{wang2018path}                 & \multicolumn{1}{c|}{0.9242}   & \multicolumn{1}{c|}{}          & \multicolumn{1}{c|}{}       & \multicolumn{1}{c|}{}                                       & \multicolumn{1}{c|}{}                                    & \multicolumn{1}{c|}{}                                    & \multicolumn{1}{c|}{}                                    & \multicolumn{1}{c|}{}                                     & \multicolumn{1}{c|}{}                                     & \multicolumn{1}{c|}{}                                  & \multicolumn{1}{c|}{}                                  & \multicolumn{1}{c|}{}                                  & \multicolumn{1}{c|}{}                                  & \multicolumn{1}{c|}{}                                   & \multicolumn{1}{c|}{}                                   & \multicolumn{1}{c|}{}                                   & \multicolumn{1}{c|}{}      & \multicolumn{1}{c|}{}                                    &        \\ \cline{2-21} 
			\multirow{-10}{*}{Amazon-Electronics} & \citep{mcauley2015inferring}         & \multicolumn{1}{c|}{0.888}    & \multicolumn{1}{c|}{}          & \multicolumn{1}{c|}{}       & \multicolumn{1}{c|}{}                                       & \multicolumn{1}{c|}{}                                    & \multicolumn{1}{c|}{}                                    & \multicolumn{1}{c|}{}                                    & \multicolumn{1}{c|}{}                                     & \multicolumn{1}{c|}{}                                     & \multicolumn{1}{c|}{}                                  & \multicolumn{1}{c|}{}                                  & \multicolumn{1}{c|}{}                                  & \multicolumn{1}{c|}{}                                  & \multicolumn{1}{c|}{}                                   & \multicolumn{1}{c|}{}                                   & \multicolumn{1}{c|}{}                                   & \multicolumn{1}{c|}{}      & \multicolumn{1}{c|}{}                                    &        \\ \hline
			& \citep{DBLP:conf/aaai/ChenHXFLSYQ23} & \multicolumn{1}{c|}{}         & \multicolumn{1}{c|}{}          & \multicolumn{1}{c|}{}       & \multicolumn{1}{c|}{0.9528}                                 & \multicolumn{1}{c|}{}                                    & \multicolumn{1}{c|}{}                                    & \multicolumn{1}{c|}{}                                    & \multicolumn{1}{c|}{}                                     & \multicolumn{1}{c|}{0.7989}                               & \multicolumn{1}{c|}{}                                  & \multicolumn{1}{c|}{}                                  & \multicolumn{1}{c|}{}                                  & \multicolumn{1}{c|}{}                                  & \multicolumn{1}{c|}{}      & \multicolumn{1}{c|}{}      & \multicolumn{1}{c|}{}      & \multicolumn{1}{c|}{}      & \multicolumn{1}{c|}{}                                    &        \\ \cline{2-21} 
			& \citep{DBLP:conf/recsys/Papso23}     & \multicolumn{1}{c|}{}         & \multicolumn{1}{c|}{}          & \multicolumn{1}{c|}{}       & \multicolumn{1}{c|}{}                                       & \multicolumn{1}{c|}{}                                    & \multicolumn{1}{c|}{}                                    & \multicolumn{1}{c|}{}                                    & \multicolumn{1}{c|}{0.4094}                               & \multicolumn{1}{c|}{}                                     & \multicolumn{1}{c|}{}                                  & \multicolumn{1}{c|}{}                                  & \multicolumn{1}{c|}{}                                  & \multicolumn{1}{c|}{}                                  & \multicolumn{1}{c|}{0.6247}                             & \multicolumn{1}{c|}{}                                   & \multicolumn{1}{c|}{}                                   & \multicolumn{1}{c|}{}      & \multicolumn{1}{c|}{}                                    &        \\ \cline{2-21} 
			& \citep{yang2022inferring}            & \multicolumn{1}{c|}{}         & \multicolumn{1}{c|}{}          & \multicolumn{1}{c|}{}       & \multicolumn{1}{c|}{}                                       & \multicolumn{1}{c|}{}                                    & \multicolumn{1}{c|}{}                                    & \multicolumn{1}{c|}{}                                    & \multicolumn{1}{c|}{}                                     & \multicolumn{1}{c|}{}                                     & \multicolumn{1}{c|}{}                                  & \multicolumn{1}{c|}{}                                  & \multicolumn{1}{c|}{}                                  & \multicolumn{1}{c|}{}                                  & \multicolumn{1}{c|}{0.75}                               & \multicolumn{1}{c|}{0.94}                               & \multicolumn{1}{c|}{0.97}                               & \multicolumn{1}{c|}{}      & \multicolumn{1}{c|}{}                                    &        \\ \cline{2-21} 
			& \citep{zhang2021learning}            & \multicolumn{1}{c|}{}         & \multicolumn{1}{c|}{}          & \multicolumn{1}{c|}{}       & \multicolumn{1}{c|}{}                                       & \multicolumn{1}{c|}{}                                    & \multicolumn{1}{c|}{}                                    & \multicolumn{1}{c|}{0.4162}                              & \multicolumn{1}{c|}{0.4542}                               & \multicolumn{1}{c|}{}                                     & \multicolumn{1}{c|}{}                                  & \multicolumn{1}{c|}{}                                  & \multicolumn{1}{c|}{}                                  & \multicolumn{1}{c|}{0.5413}                            & \multicolumn{1}{c|}{0.6588}                             & \multicolumn{1}{c|}{}                                   & \multicolumn{1}{c|}{}                                   & \multicolumn{1}{c|}{}      & \multicolumn{1}{c|}{}                                    &        \\ \cline{2-21} 
			& \citep{zhang2019identifying}         & \multicolumn{1}{c|}{0.9111}   & \multicolumn{1}{c|}{}          & \multicolumn{1}{c|}{}       & \multicolumn{1}{c|}{}                                       & \multicolumn{1}{c|}{}                                    & \multicolumn{1}{c|}{}                                    & \multicolumn{1}{c|}{}                                    & \multicolumn{1}{c|}{}                                     & \multicolumn{1}{c|}{}                                     & \multicolumn{1}{c|}{}                                  & \multicolumn{1}{c|}{}                                  & \multicolumn{1}{c|}{}                                  & \multicolumn{1}{c|}{}                                  & \multicolumn{1}{c|}{}                                   & \multicolumn{1}{c|}{}                                   & \multicolumn{1}{c|}{}                                   & \multicolumn{1}{c|}{}      & \multicolumn{1}{c|}{}                                    & 0.9076 \\ \cline{2-21} 
			\multirow{-6}{*}{Amazon-Cellphones}   & \citep{wang2018path}                 & \multicolumn{1}{c|}{0.9306}   & \multicolumn{1}{c|}{}          & \multicolumn{1}{c|}{}       & \multicolumn{1}{c|}{}                                       & \multicolumn{1}{c|}{}                                    & \multicolumn{1}{c|}{}                                    & \multicolumn{1}{c|}{}                                    & \multicolumn{1}{c|}{}                                     & \multicolumn{1}{c|}{}                                     & \multicolumn{1}{c|}{}                                  & \multicolumn{1}{c|}{}                                  & \multicolumn{1}{c|}{}                                  & \multicolumn{1}{c|}{}                                  & \multicolumn{1}{c|}{}                                   & \multicolumn{1}{c|}{}                                   & \multicolumn{1}{c|}{}                                   & \multicolumn{1}{c|}{}      & \multicolumn{1}{c|}{}                                    &        \\ \hline
			& \citep{DBLP:conf/aaai/ChenHXFLSYQ23} & \multicolumn{1}{c|}{}         & \multicolumn{1}{c|}{}          & \multicolumn{1}{c|}{}       & \multicolumn{1}{c|}{0.9643}                                 & \multicolumn{1}{c|}{}                                    & \multicolumn{1}{c|}{}                                    & \multicolumn{1}{c|}{}                                    & \multicolumn{1}{c|}{}                                     & \multicolumn{1}{c|}{0.7866}                               & \multicolumn{1}{c|}{}                                  & \multicolumn{1}{c|}{}                                  & \multicolumn{1}{c|}{}                                  & \multicolumn{1}{c|}{}                                  & \multicolumn{1}{c|}{}      & \multicolumn{1}{c|}{}      & \multicolumn{1}{c|}{}      & \multicolumn{1}{c|}{}      & \multicolumn{1}{c|}{}                                    &        \\ \cline{2-21} 
			& \citep{DBLP:conf/www/BibasSJ23}                & \multicolumn{1}{c|}{}         & \multicolumn{1}{c|}{}          & \multicolumn{1}{c|}{}       & \multicolumn{1}{c|}{}                                       & \multicolumn{1}{c|}{}                                    & \multicolumn{1}{c|}{}                                    & \multicolumn{1}{c|}{}                                    & \multicolumn{1}{c|}{}                                     & \multicolumn{1}{c|}{}                                     & \multicolumn{1}{c|}{0.316}                             & \multicolumn{1}{c|}{0.092}                             & \multicolumn{1}{c|}{}                                  & \multicolumn{1}{c|}{0.233}                             & \multicolumn{1}{c|}{ 0.332} & \multicolumn{1}{c|}{}      & \multicolumn{1}{c|}{}      & \multicolumn{1}{c|}{}      & \multicolumn{1}{c|}{}                                    &        \\ \cline{2-21} 
			& \citep{liu2021item}                  & \multicolumn{1}{c|}{0.7417}   & \multicolumn{1}{c|}{0.7408}    & \multicolumn{1}{c|}{0.8805} & \multicolumn{1}{c|}{}                                       & \multicolumn{1}{c|}{0.7049}                              & \multicolumn{1}{c|}{}                                    & \multicolumn{1}{c|}{}                                    & \multicolumn{1}{c|}{}                                     & \multicolumn{1}{c|}{}                                     & \multicolumn{1}{c|}{}                                  & \multicolumn{1}{c|}{}                                  & \multicolumn{1}{c|}{}                                  & \multicolumn{1}{c|}{}                                  & \multicolumn{1}{c|}{}                                   & \multicolumn{1}{c|}{}                                   & \multicolumn{1}{c|}{}                                   & \multicolumn{1}{c|}{}      & \multicolumn{1}{c|}{}                                    &        \\ \cline{2-21} 
			& \citep{liu2020decoupled}             & \multicolumn{1}{c|}{}         & \multicolumn{1}{c|}{}          & \multicolumn{1}{c|}{}       & \multicolumn{1}{c|}{}                                       & \multicolumn{1}{c|}{}                                    & \multicolumn{1}{c|}{}                                    & \multicolumn{1}{c|}{}                                    & \multicolumn{1}{c|}{0.45}                                 & \multicolumn{1}{c|}{}                                     & \multicolumn{1}{c|}{}                                  & \multicolumn{1}{c|}{}                                  & \multicolumn{1}{c|}{}                                  & \multicolumn{1}{c|}{}                                  & \multicolumn{1}{c|}{0.6}                                & \multicolumn{1}{c|}{}                                   & \multicolumn{1}{c|}{}                                   & \multicolumn{1}{c|}{}      & \multicolumn{1}{c|}{0.327}                               &        \\ \cline{2-21} 
			\multirow{-5}{*}{Amazon-Clothing}     & \citep{zhao2017deep}                 & \multicolumn{1}{c|}{}         & \multicolumn{1}{c|}{}          & \multicolumn{1}{c|}{0.983}  & \multicolumn{1}{c|}{}                                       & \multicolumn{1}{c|}{}                                    & \multicolumn{1}{c|}{}                                    & \multicolumn{1}{c|}{}                                    & \multicolumn{1}{c|}{}                                     & \multicolumn{1}{c|}{}                                     & \multicolumn{1}{c|}{}                                  & \multicolumn{1}{c|}{}                                  & \multicolumn{1}{c|}{}                                  & \multicolumn{1}{c|}{}                                  & \multicolumn{1}{c|}{}                                   & \multicolumn{1}{c|}{}                                   & \multicolumn{1}{c|}{}                                   & \multicolumn{1}{c|}{}      & \multicolumn{1}{c|}{}                                    &        \\ \hline
			& \citep{DBLP:conf/recsys/Papso23}     & \multicolumn{1}{c|}{}         & \multicolumn{1}{c|}{}          & \multicolumn{1}{c|}{}       & \multicolumn{1}{c|}{}                                       & \multicolumn{1}{c|}{}                                    & \multicolumn{1}{c|}{}                                    & \multicolumn{1}{c|}{}                                    & \multicolumn{1}{c|}{0.5659}                               & \multicolumn{1}{c|}{}                                     & \multicolumn{1}{c|}{}                                  & \multicolumn{1}{c|}{}                                  & \multicolumn{1}{c|}{}                                  & \multicolumn{1}{c|}{}                                  & \multicolumn{1}{c|}{0.7697}                             & \multicolumn{1}{c|}{}                                   & \multicolumn{1}{c|}{}                                   & \multicolumn{1}{c|}{}      & \multicolumn{1}{c|}{}                                    &        \\ \cline{2-21} 
			& \citep{wu2022towards}                & \multicolumn{1}{c|}{}         & \multicolumn{1}{c|}{}          & \multicolumn{1}{c|}{}       & \multicolumn{1}{c|}{}                                       & \multicolumn{1}{c|}{}                                    & \multicolumn{1}{c|}{0.485}                               & \multicolumn{1}{c|}{}                                    & \multicolumn{1}{c|}{}                                     & \multicolumn{1}{c|}{}                                     & \multicolumn{1}{c|}{}                                  & \multicolumn{1}{c|}{}                                  & \multicolumn{1}{c|}{0.641}                             & \multicolumn{1}{c|}{}                                  & \multicolumn{1}{c|}{}                                   & \multicolumn{1}{c|}{}                                   & \multicolumn{1}{c|}{}                                   & \multicolumn{1}{c|}{0.495} & \multicolumn{1}{c|}{}                                    &        \\ \cline{2-21} 
			& \citep{liu2021item}                  & \multicolumn{1}{c|}{0.8129}   & \multicolumn{1}{c|}{0.7977}    & \multicolumn{1}{c|}{0.9505} & \multicolumn{1}{c|}{}                                       & \multicolumn{1}{c|}{0.7546}                              & \multicolumn{1}{c|}{}                                    & \multicolumn{1}{c|}{}                                    & \multicolumn{1}{c|}{}                                     & \multicolumn{1}{c|}{}                                     & \multicolumn{1}{c|}{}                                  & \multicolumn{1}{c|}{}                                  & \multicolumn{1}{c|}{}                                  & \multicolumn{1}{c|}{}                                  & \multicolumn{1}{c|}{}                                   & \multicolumn{1}{c|}{}                                   & \multicolumn{1}{c|}{}                                   & \multicolumn{1}{c|}{}      & \multicolumn{1}{c|}{}                                    &        \\ \cline{2-21} 
			& \citep{rakesh2019linked}             & \multicolumn{1}{c|}{0.9347}   & \multicolumn{1}{c|}{}          & \multicolumn{1}{c|}{}       & \multicolumn{1}{c|}{}                                       & \multicolumn{1}{c|}{}                                    & \multicolumn{1}{c|}{}                                    & \multicolumn{1}{c|}{}                                    & \multicolumn{1}{c|}{}                                     & \multicolumn{1}{c|}{}                                     & \multicolumn{1}{c|}{}                                  & \multicolumn{1}{c|}{}                                  & \multicolumn{1}{c|}{}                                  & \multicolumn{1}{c|}{}                                  & \multicolumn{1}{c|}{}                                   & \multicolumn{1}{c|}{}                                   & \multicolumn{1}{c|}{}                                   & \multicolumn{1}{c|}{}      & \multicolumn{1}{c|}{}                                    &        \\ \cline{2-21} 
			\multirow{-5}{*}{Amazon-Music}        & \citep{mcauley2015inferring}         & \multicolumn{1}{c|}{0.9043}   & \multicolumn{1}{c|}{}          & \multicolumn{1}{c|}{}       & \multicolumn{1}{c|}{}                                       & \multicolumn{1}{c|}{}                                    & \multicolumn{1}{c|}{}                                    & \multicolumn{1}{c|}{}                                    & \multicolumn{1}{c|}{}                                     & \multicolumn{1}{c|}{}                                     & \multicolumn{1}{c|}{}                                  & \multicolumn{1}{c|}{}                                  & \multicolumn{1}{c|}{}                                  & \multicolumn{1}{c|}{}                                  & \multicolumn{1}{c|}{}                                   & \multicolumn{1}{c|}{}                                   & \multicolumn{1}{c|}{}                                   & \multicolumn{1}{c|}{}      & \multicolumn{1}{c|}{}                                    &        \\ \hline
			& \citep{yang2022inferring}            & \multicolumn{1}{c|}{}         & \multicolumn{1}{c|}{}          & \multicolumn{1}{c|}{}       & \multicolumn{1}{c|}{}                                       & \multicolumn{1}{c|}{}                                    & \multicolumn{1}{c|}{}                                    & \multicolumn{1}{c|}{}                                    & \multicolumn{1}{c|}{}                                     & \multicolumn{1}{c|}{}                                     & \multicolumn{1}{c|}{}                                  & \multicolumn{1}{c|}{}                                  & \multicolumn{1}{c|}{}                                  & \multicolumn{1}{c|}{}                                  & \multicolumn{1}{c|}{0.68}                               & \multicolumn{1}{c|}{0.86}                               & \multicolumn{1}{c|}{0.9}                                & \multicolumn{1}{c|}{}      & \multicolumn{1}{c|}{}                                    &        \\ \cline{2-21} 
			& \citep{zhang2021learning}            & \multicolumn{1}{c|}{}         & \multicolumn{1}{c|}{}          & \multicolumn{1}{c|}{}       & \multicolumn{1}{c|}{}                                       & \multicolumn{1}{c|}{}                                    & \multicolumn{1}{c|}{}                                    & \multicolumn{1}{c|}{0.2504}                              & \multicolumn{1}{c|}{0.292}                                & \multicolumn{1}{c|}{}                                     & \multicolumn{1}{c|}{}                                  & \multicolumn{1}{c|}{}                                  & \multicolumn{1}{c|}{}                                  & \multicolumn{1}{c|}{0.3569}                            & \multicolumn{1}{c|}{0.4857}                             & \multicolumn{1}{c|}{}                                   & \multicolumn{1}{c|}{}                                   & \multicolumn{1}{c|}{}      & \multicolumn{1}{c|}{}                                    &        \\ \cline{2-21} 
			\multirow{-3}{*}{Amazon-Baby}         & \citep{zhang2019identifying}         & \multicolumn{1}{c|}{0.8777}   & \multicolumn{1}{c|}{}          & \multicolumn{1}{c|}{}       & \multicolumn{1}{c|}{}                                       & \multicolumn{1}{c|}{}                                    & \multicolumn{1}{c|}{}                                    & \multicolumn{1}{c|}{}                                    & \multicolumn{1}{c|}{}                                     & \multicolumn{1}{c|}{}                                     & \multicolumn{1}{c|}{}                                  & \multicolumn{1}{c|}{}                                  & \multicolumn{1}{c|}{}                                  & \multicolumn{1}{c|}{}                                  & \multicolumn{1}{c|}{}                                   & \multicolumn{1}{c|}{}                                   & \multicolumn{1}{c|}{}                                   & \multicolumn{1}{c|}{}      & \multicolumn{1}{c|}{}                                    & 0.8365 \\ \hline
		\end{tabular}%
	}
\end{table*}
As shown in Table \ref{tab:my-table1}, the commonly used evaluation indicators for complementary recommendations are as follows:
\newline \indent \textbf{Precision.} The ratio of the number of correct products recommended by the recommendation system to the number of products actually recommended to the user. Only \citep{liu2021item} employs precision as the evaluation metric.
\newline \indent \textbf{Accuracy.} The ratio of correctly predicted samples to the total samples. It can be observed that for different datasets, the accuracy of the method described in reference \citep{wang2018path} is consistently the highest.
\newline \indent \textbf{AUC.} The area under the ROC curve reflects the ranking ability of the model. In the clothing domain, the AUC value of \citep{zhao2017deep} is the highest, reaching 0.983.
\newline \indent \textbf{Recall.} The ratio of the number of correct products recommended by the recommendation system to the number of products actually clicked by the user. The recall rate of reference \citep{DBLP:conf/aaai/ChenHXFLSYQ23} is the highest.
\newline \indent \textbf{NDCG.}  Cares about whether the found products are placed in a more conspicuous position in the user interface, emphasizing the "sequence" of prediction.
\newline \indent \textbf{ HR.} Cares about whether what users want is recommended, and emphasizes the "accuracy" of predictions.
\section{Future research and exploration}
After reviewing existing complementary recommendation methods based on e-commerce platforms, we propose the following future research directions to promote the development of this field.
\subsection{Modeling of complementary relationships}
Currently, the complementarity relationships between products are mostly modeled based on bought together data, which is a single and fundamentally flawed approach, as products bought together may not necessarily be complementary. Therefore, we propose three alternative modeling approaches for consideration in future research.
\subsubsection{Modeling from a semantic perspective}
\
\newline \indent \textbf{E-commerce field.}
\begin{itemize}
	\item Fashion category: Existing fashion category complementary relationship modeling is mainly based on pictures and text descriptions of clothing. However, pictures may not be able to show the tailoring, texture and fitting effect of clothing, and text descriptions are also difficult to convey how the clothing is actually worn effect. This makes it difficult for users to judge whether the clothing is suitable for them. Therefore, 3D virtual fitting technology can be used to allow users to try on clothes in a virtual environment, and recommend suitable clothing styles, colors and matching methods based on the user's personal style, preferences and physical characteristics.
	\item Diet category: Currently, the modeling of complementary relationships in diet categories only relies on user co-purchase data. For example, users often buy soy milk and fried dough sticks at the same time during breakfast time, while ignoring the complementary relationship between ingredients between foods. For example, iron is an essential mineral for the human body, and vitamin C helps the body absorb iron. Therefore, combining iron-rich foods (such as meat, beans, green leafy vegetables, etc.) with vitamin C-rich foods (such as citrus fruits, tomatoes, red peppers, etc.) can improve the absorption efficiency of iron. Based on the complementary relationship between ingredients in food, we can help us obtain a more comprehensive and balanced nutritional intake, and improve the taste and flavor of food.
	\item Books Category: Books can be judged as complementary based on the topics they cover and the content they explain. For example, a book that introduces historical events can complement a book about the biography of historical figures. Books about historical events provide an understanding of the context of historical development, important events, and the impact on society and culture, and books about the biographies of historical figures provide insights into the roles, thoughts, behaviors, and influences of individuals in history. This helps readers gain more comprehensive historical knowledge.
\end{itemize}
\indent \indent \textbf{Non-e-commerce field.}
\begin{itemize}
	\item Movie recommendation: Most of the existing movie recommendations are based on the user's favorite actors, genres, etc., making the recommended movie types too single. Therefore, complementary recommendations can be made based on the theme and storyline of the movie. For example, two movies may have common themes but portray them from different angles or perspectives. In this case, the two movies are recommended to the user as a complementary relationship, which can provide users with a more comprehensive and comprehensive movie-watching experience.
	\item News push: Judge complementary relationships based on the viewpoints of content in news reports, and push news reports with contrary or controversial viewpoints to users' reading preferences. This can break the barriers of information filtering and avoid falling into information homogeneity and bias. Help users understand more opinions and positions, as well as the views of different social groups, promote thinking and debate, and provide more comprehensive information.
\end{itemize}
\subsubsection{Modeling from a usage perspective}
\
\newline \indent Current complementary recommendation systems usually make recommendations based on product propertiess and user behavior, but rarely take into account the needs of the current scene and situation for the use of the product. However, making recommendations based on the user's current scenario and context can provide more personalized and relevant suggestions. First, the recommendation system can determine the current scene based on the user's location, weather, time and other factors, and provide relevant product recommendations to the user. For example, on a hot summer day, refreshing drinks, sunscreen, and cool clothing can be recommended to help users stay comfortable in hot environments. If the user plans to travel to Harbin in winter, products such as warm scarves, thick hats, snow boots, etc. can be recommended to the user based on Harbin's weather conditions to meet the user's need for warmth in a cold environment.
\newline \indent Secondly, situation-based recommendations can span different product categories, provide comprehensive suggestions, and meet the multiple needs of users. The recommendation system can analyze the user's activities, interests and hobbies and other information, and recommend relevant products based on the current situation. For example, if a user is arranging a family party, the recommendation system can recommend food, drinks, decorations, entertainment toys and other related products across categories to help the user create a complete and rich party experience. This recommendation provides more comprehensive and diverse suggestions, allowing users to easily find products that meet different aspects of their needs.
\newline \indent Through recommendations based on users' current scenarios and situations, complementary recommendation systems can better meet users' actual needs and provide more personalized and appropriate product recommendations. This method can not only enhance the user's shopping experience, but also improve the effectiveness of the recommendation system and user satisfaction.
\subsubsection{Modeling from a quantitative perspective}
\
\newline \indent \textbf{One-to-one complementation.} Current one-to-one complementary relationship recommendation methods usually focus on the direct link between two items, ignoring the existence of complex paths. However, accounting for complex paths can further increase the number of complementary products, thus providing more choices and possibilities. For example, product A is considered a substitute for product B, and product B is considered a complement to product C. If we consider complex paths, we can infer that product A is also a complement to product C. By inferring the relationship between indirectly linked products, a more detailed and comprehensive complementary relationship between products can be established.
\newline \indent \textbf{Combination complementation.}Currently, most methods of complementary relationship modeling focus on one-to-one complementary relationships, but sometimes this simple one-to-one complementarity cannot fully meet the needs of users. Taking the user's purchase of a cellphone as an example, the traditional one-to-one complementary relationship modeling method may only regard the charging head or data cable as a complementary product of the cellphone. However, for users, the function of charging their cellphones can only be realized if they have both a charging head and a data cable. Therefore, the combined complementary relationship modeling method takes this into account and treats the combination of charging head and data cable as a more comprehensive complementary combination to meet the actual needs of users.
\subsection{Sequential complementary recommendations}
The current focus of complementary recommendation mainly lies in modeling the complementarity between products. However, this approach may result in recommended products lacking personalization, with the same items being recommended to every user despite their individual preferences. Therefore, by combining complementary recommendation with sequential recommendation and considering users' purchasing interests, we can enhance recommendation diversity and increase user satisfaction.
\subsubsection{Sequential user intent recognition}
\
\newline \indent Traditional complementary recommendation algorithms are mainly based on users' historical behaviors and preferences, but often ignore the importance of time dimension. However, the relationship between the user and the product changes over time. For example, users prefer cost-effective products when they are students, and are more inclined to high-quality products after graduation. Future developments will focus on temporal modeling of user behavior, taking into account user interest evolution, cyclical needs, and temporal correlation to more accurately predict user interests and needs. The recommendation algorithm can provide suitable recommendation content in different time periods according to the user's time limit and priority. In order for users to be satisfied with the recommended product, the system should be updated repeatedly with new information. At present, complementary recommendation is an under-explored direction in the fusion of temporal information, which is worthy of further study. Mainly divided into the following aspects:
\newline \indent \textbf{Multi-dimensional time information fusion.} It is possible to consider the integration of multiple dimensions of time information, such as the time interval of user behavior, the release time of products, and the frequency of updates, in order to more comprehensively understand the evolution of user interests.
\newline \indent \textbf{The balance of long-term and short-term interests.} Users' interests can include long-term interests and short-term interests, and the recommendation algorithm needs to be able to accurately distinguish and model these two interests according to time information, and generate targeted recommendation results accordingly. This will help improve the personalization of recommendations and user satisfaction. Different time scales have different effects on recommendation effectiveness. How to choose the appropriate time scale so that the time information can better reflect the evolution of users' interests is a problem that needs to be studied. In addition, how to ensure the accuracy of recommendation while improving the real-time and efficiency of recommendation algorithm is also a difficult problem that needs to be balanced.
\newline \indent \textbf{User intention recognition in dialogue and interaction scenarios.} At present, most of the interaction scenarios are based on the user's final intention as the user's real intention, but when the user's dialogue has a transition or a number of complex intentions, the system can not well identify the user's real intention. For example, a user: "What's the weather like today? I want to buy a swimsuit to go swimming." Obviously, whether the user buys a swimsuit and what kind of swimsuit is related to today's weather, so the system should first respond to the user's weather query intent, and then recommend him a suitable swimsuit.
\subsubsection{Cold start and scene migration issues in sequence complementary recommendation}
\
\newline \indent The cold-start problem refers to the lack of sufficient historical data or information, which makes it difficult for the recommendation system to accurately recommend. According to the lack of data, cold-start problems are mainly divided into two categories:
\newline \indent \textbf{The user cold-starts.} This type of cold-start problem refers to the challenge of facing new users. Due to the lack of personal historical behavior data of new users, it is difficult for the recommendation system to accurately understand their interests and preferences, so that personalized recommendations cannot be made. The methods to solve the problem of cold-start of users include obtaining the characteristics of new users based on registration information, user attributes, social networks, etc., and using content-based recommendation, popular recommendation and other strategies to carry out preliminary personalized recommendation.
\newline \indent \textbf{Cold-start of products.} When new products are introduced into the recommendation system, these new products may lack sufficient historical interaction data, and it is difficult for the recommendation system to understand their characteristics, categories and matching degree with users, which makes it difficult for the recommendation system to recommend new products to users who are interested in them. Solutions to the product cold-start problem include using the content information of the new product, attribute characteristics, and using other behavioral data of the user (such as purchase history, ratings, browsing, etc.) to infer the match between the new product and the user.
\newline \indent When the system faces new users or new products, it can also use the method of transfer learning to transfer the knowledge of the existing scene to the new scene. This can be done by collecting some initial data in a new scenario and then using domain adaptive techniques to adjust the model parameters to the characteristics of the new scenario. With the accumulation of data in the new scene, the model can be updated incrementally by using the incremental learning method to adapt to the changes of the new scene. This allows new data to be used to fine-tune the model and improve recommendation performance. When training the model, the data of multiple scenarios can be considered at the same time to carry out multi-task learning. This allows you to share feature representations, extract shared and specific features, and make recommendations across multiple scenarios.
\subsubsection{Diversification, personalization and timing of complementary recommendations}
\
\newline \indent In complementary recommendations, diversification, personalization and timing are three important considerations.
\newline \indent \textbf{Diversified recommendation.} Complementary recommendations focus on providing diverse recommendations. It not only focuses on the user's main preferences, but also recommends some items related to the user's interests but are different, to expand the diversity of recommendations and help users discover new and potential areas of interest. In this way, the recommendation system can avoid the situation of excessive preference for one type of products, so as to provide more rich and interesting recommendation choices. At present, there are few researches on complementary recommendation of diversity, which can be used as the future development direction.
\newline \indent \textbf{Personalized recommendation.} Personalized recommendation needs to provide unique recommendation results for each user based on their personal preferences and preferences. Machine learning algorithms can be used to build user models and make personalized recommendations based on information such as historical behavior, interest labels, and so on. Recommendations can also be dynamically adjusted based on user feedback and preferences to adapt to user changes.
\newline \indent \textbf{Sequential recommendation.} Sequential recommendation considers the timeliness and real-time of recommendation results. This is important for time-sensitive content such as news, events, specials, etc. The recommendation system can update the recommendation results in time according to the user's real-time behavior, the current environment and the latest information, and present the most relevant and latest content to the user. This can increase the user's dependency and satisfaction with the recommendation system.
\subsection{A new paradigm for recommendation systems}
The emergence of large-scale model technology has changed the research and application paradigm in the field of artificial intelligence. Therefore, combining complementary recommendation with large models can achieve better recommendation results.
\subsubsection{When the recommendation system encounters a large language model}
\
\newline \indent \textbf{Better semantic understanding.} The powerful semantic understanding of large language models is indeed a key advantage in recommendation systems. Through in-depth context understanding and semantic information extraction, large language models can better understand users' queries, needs and preferences, so as to provide more accurate, relevant and personalized recommendation results. Specifically, large language models can capture the associations, logic, and grammatical structures between words by analyzing the context of text. This allows it to better understand the meaning of the text, not just the surface literal meaning. In addition, the large language model can extract the implicit semantic information in the text. This means that it can find hidden associations and meanings in text, even those that are not explicitly expressed. By capturing these implicit semantics, large language models can better understand user preferences, interests, and preferences to provide more accurate and personalized recommendations.
\newline \indent \textbf{Generate textual explanations.} When a recommendation system recommends an item to a user, large language models can come into play, generating text to explain the reasons and advantages behind the recommendation results. By generating a text that describes the features and benefits of the item, the model can explain to the user why the product is a good choice, helping the user to become more aware of the recommended product. This helps users make informed purchasing decisions and enhances trust in the recommendation system. At the same time, this explanatory text also provides users with a better shopping experience, making them more satisfied and confident in accepting the recommendation results.
\subsubsection{Generative retrieval recommendation system}
\
\newline \indent In traditional recommendation systems, recall and ranking strategies are usually adopted. The recall phase selects a set of candidates from a large number of items and then sorts these candidates using a sorting model. However, there are two limitations to this recommendation paradigm: 1?The recalled items may not meet the needs of users. During the recall phase, the selected candidate set may not fully match the individual needs of the user. This can result in recommendations that do not match the user's expectations. 2?Traditional feedback methods are inefficient. Traditional recommendation systems usually rely on passive feedback from users, such as clicking behavior, which is sometimes not accurate and timely enough to fully express the real needs and preferences of users.
\newline \indent To overcome these limitations, more and more research and applications have begun to adopt generative retrieval models, such as ChatGPT and other large language models. Different from traditional query candidate matching methods, generative models can generate personalized product or recommendation results according to the specific needs of users. With natural language instructions, users can more accurately express requirements, and the model can generate recommendations based on these instructions that meet the user's expectations. Moreover, the generative retrieval model has certain creativity and flexibility, and can generate novel and personalized recommendation results. These results may not be considered by traditional recommendation systems, which can better meet the specific needs of users.
\newline \indent Despite its potential in recommendations, generative retrieval models also need to address challenges such as diversity and quality control of generated results. With the continuous development and improvement of large language models, the application of generative retrieval model in recommendation system is very broad.
\subsubsection{Multi-interest recall in complementary recommendation systems}
\
\newline \indent The core of multi-interest recall is to capture multiple interest expressions of users through historical behavior, and use different interest expressions to carry out differentiated parallel recall of multi-way interest perception. There are two major challenges: negative example sampling strategy and interest route collapse.
\newline \indent \textbf{Hard negative sampling strategies.} The existing multi-interest recall training method will make common negative example sampling strategies ineffective. Therefore, hard negative sampling strategies can be used. An ideal hard negative sampling strategies should be to sample samples that are easily misclassified by the head of specific interest, and the difficulty of negative examples should be adjustable to adapt to different data distributions. Drawing on the idea of contrastive learning, we adopt real negative examples that are highly similar to the representation of interest.
\newline \indent \textbf{Routing Regularization.} It has been observed that after multiple rounds of training, interest tends to focus excessively on a single item in a sequence of behaviors. In this case, only a small part of the user's historical data is taken into account, resulting in impaired expressiveness of multi-interest representations. Therefore, the routing regularization strategy is used to constrain the variance of the routing matrix.
\subsection{Recommendation system under complex data}
Currently, complementary recommendation primarily utilizes publicly available data from Amazon. However, in practical applications, more complex datasets may arise. Below are the methods for dealing with different data.
\subsubsection{Multimodal data}
\
\newline \indent Most of the traditional complementary recommendation methods use the text and image information of products, and the future complementary recommendation algorithm of products can consider the integration of multi-modal data such as audio and video for recommendation. This will enable the recommendation system to better adapt to the diverse needs of users and provide more attractive recommendation results.
\newline \indent \textbf{Image recognition and recommendation.} Through image recognition technology, products and objects can be identified according to the pictures provided by the user or the scene images captured by the camera, so as to provide users with product recommendations related to the objects in the image.For example, if a user takes a picture of a kitchen, the system can identify products in the kitchen and recommend complementary products accordingly.
\newline \indent \textbf{Voice interaction and recommendation.} Voice interaction will make product recommendations more intelligent and personalized. Through the user's voice command, we can understand the user's needs and preferences, and provide corresponding product recommendations. For example, users can ask for a specific brand's product or describe their purchase intention by voice, and the system can make complementary product recommendations based on the voice content.
\newline \indent \textbf{3D virtual fitting.} The traditional clothing recommendation is only based on the clothing matching of most users, without considering the preferences and body types of each user. 3D virtual fitting allows users to have a fitting experience in a virtual environment and view recommended clothing styles. Users can rotate and adjust the Angle of view to see how the garment looks on their body. Users can also provide feedback, such as comments on style, color, size, etc., to help the system further optimize the recommendation results.
\subsubsection{There is no labeled data for large scale}
\
\newline \indent In the absence of clear labels or ratings, making recommendations with massive amounts of unlabeled data is a challenge. In this case, you can use the following methods:
\newline \indent \textbf{Transfer learning and pre-trained models.} Leverage pre-trained models, such as the Transformer model in deep learning, to learn generic representational and semantic information from large-scale unlabeled data. These pre-trained models can capture underlying structures and patterns in the data and apply these representations to recommendation tasks. Through transfer learning, existing knowledge can be transferred to new recommendation tasks to improve the performance of the model.
\newline \indent \textbf{Reinforcement learning.} Reinforcement learning technology is used to design a recommendation system to interact with users and optimize recommendation strategies through trial and error and feedback. In this approach, the recommendation system can interact with the user and observe the user's behavior and feedback. By collecting this feedback data, reinforcement learning algorithms can gradually optimize recommendation strategies to maximize user satisfaction. This approach can gradually learn and optimize recommendations through interaction with users without annotated data.
\subsubsection{Small sample, zero sample data}
\
\newline \indent The small-sample and zero-sample data recommendation system is capable of effective recommendation in the case of scarce data. In this case, the recommendation system needs to make full use of limited data and other supporting information to understand the relationship between user interests and products. Here are some common techniques:
\newline \indent \textbf{Knowledge graph and external data.} Knowledge graph and external data. The use of external knowledge graphs or other publicly available data sources can complement small or zero-sample data. This data can provide attributes, relationships, or related domain expertise to enhance the understanding of the recommendation system.
\newline \indent \textbf{User interaction and feedback.} In the case of small and zero sample data, user interaction and feedback become especially important. The system can actively interact with users and encourage users to provide feedback or evaluation to obtain more information to improve the recommendation effect. This can be done through user surveys, ratings, reviews, or real-time feedback.
\newline \indent \textbf{Generative models and adversarial learning.} The generation model can increase the amount of data by generating new samples, thus solving the problem of small sample and zero sample data. By means of adversarial learning, the generative model can be trained to generate samples similar to real data and expand the data set of the recommendation system.
\section{CONCLUSION}
\label{sec:Conclusions}
This review systematically reviewed the research status and progress of complementary recommendation in the field of e-commerce. Firstly, we systematically reviewed the methods of modeling complementary relationships between products, which laid an important foundation for in-depth research on complementary recommendation. Secondly, we analyzed existing complementary recommendation models, enabling readers to fully understand the research status in this field. Finally, we proposed future development directions, hoping to promote the advancement of complementary recommendation. In summary, complementary recommendation has broad application prospects and research value. Future research should continue to deepen the exploration of theories and methods of complementary recommendation, promoting its widespread application in practice.
\section*{Acknowledgments}
This work is supported by the National Science Foundation of China (No.62162048,62262047), the Natural Science Foundation of Inner Mongolia in China(No. 2021MS01023), the Self-topic Project of Engineering Research Center of Ecological Big Data, Ministry of Education, and Inner Mongolia Science and Technology Plan
Project (2021GG0164).
\bibliographystyle{elsarticle-num} 
\bibliography{sample-base}

\begin{landscape}
\includepdf[pages={1},angle=90,]{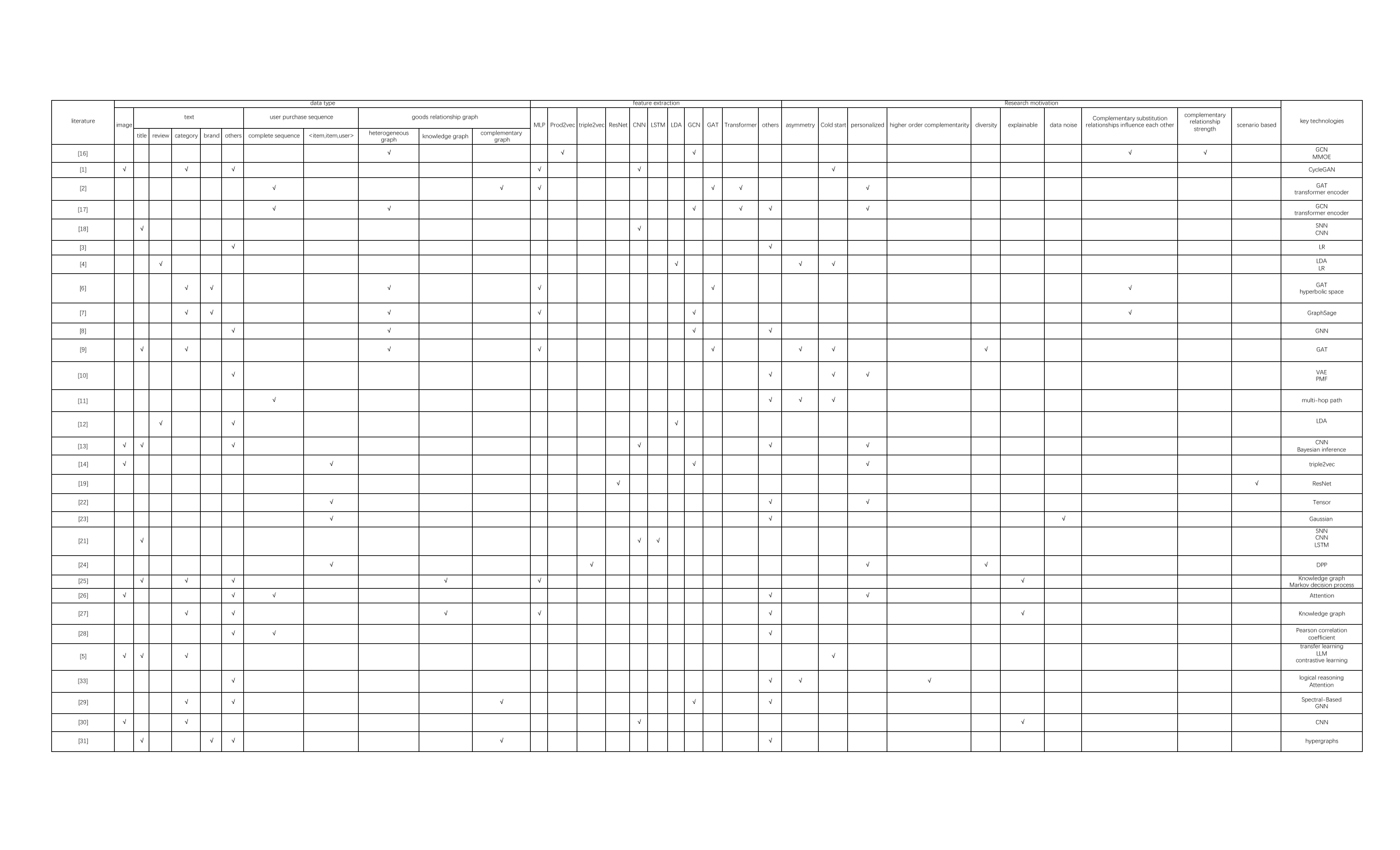}
\end{landscape}

\end{document}